\documentclass[11pt]{article} 

\usepackage{epsf,natbib} 
\usepackage{amssymb}
\usepackage{amsmath}
\usepackage{graphicx}
\usepackage{subfigure}
\usepackage{makeidx}
\usepackage{multicol}
\usepackage{color}
\usepackage{enumerate} 

\begin{document}
\title{The Tangent Exponential Model } 
\author{
A.~C.~Davison and N.~Reid
\footnote{Anthony Davison is Professor of Statistics, Institute of Mathematics, Ecole Polytechnique F\'ed\'erale de Lausanne, EPFL-FSB-MATH-STAT, Station 8, 1015 Lausanne, Switzerland ({\tt Anthony.Davison@epfl.ch}).  Nancy Reid is University Professor of Statistical Sciences, Department of Statistical Sciences, University of Toronto, 700 University Ave, 9th Floor, Toronto, Ontario M5G 1Z5, Canada ({\tt  nancym.reid@utoronto.ca}).} 
}
\date{\today}

\maketitle {\narrower\smallskip\centerline{\bf Summary} \noindent  

The likelihood function is central to  both frequentist and Bayesian formulations of parametric statistical inference, and large-sample approximations to the sampling distributions of estimators and test statistics, and to posterior densities, are widely used in practice.  Improved approximations have been widely studied and can provide highly accurate inferences when samples are small or there are many nuisance parameters.  This article reviews improved approximations based on the tangent exponential model developed in a series of articles by D.~A.~S.~Fraser and co-workers, attempting to explain the theoretical basis of this model and to provide a guide to the associated literature, including a partially-annotated bibliography.
}


\maketitle 


%
%




\def\half{{\textstyle{1\over 2}}}
\def\third{{\textstyle{1\over 3}}}
\def\quarter{{\textstyle{1\over 4}}}
\def\sixth{{\textstyle{1\over 6}}}
\def\frac#1#2{{\textstyle{#1\over#2}}}
\def\Beta{{\rm B}}
\def\dif{{\mathrm{d}}}
\def\argmin{{\rm argmin}}  
\def\diag{{\rm diag}}  
\def\mod{{\rm mod}}  
\def\tr{{\rm tr}}
\def\sign{{\rm sign}}
\def\ow{{\rm otherwise}}
\def\rest{{\rm rest}}  
\def\obs{{\rm obs}}  
\def\endex{{\hfill $\square$\\ \smallskip}}
\def\redbf#1{{\color{red}\bf #1}}%
\def\red#1{{\color{red} #1}}%
\def\white#1{{\color{black} #1}}%

\def\source#1{{\tiny(Source: #1)}}

\newcommand{\D}[1]{\mathrm{d}{#1}}
\newcommand{\sD}[1]{\,\mathrm{d}{#1}}

\def\dfres{{{\rm df}_{\rm res}}}


\def\b#1{\textbf{\color{red} #1}}%
\def\bfn#1{\footnote{\textbf{\color{red} #1}}}%

\def\Natural{\mathbb{N}}
\def\Integer{\mathbb{Z}}
\def\Real{\mathbb{R}}
\def\Rational{\mathbb{Q}}
\def\Imaginary{\mathbb{I}}
\def\Complex{\mathbb{C}}
\def\Disk{\mathbb{D}}
\def\bbE{\mathbb{E}}
\def\bbQ{\mathbb{Q}}

\def\bi{\begin{itemize}}
\def\ei{\end{itemize}}
\def\bd{\begin{description}}
\def\ed{\end{description}}
\def\ben{\begin{enumerate}}
\def\een{\end{enumerate}}
\def\bv{\begin{verbatim}}
\def\ev\end{verbatim}

\def\bth{{$b{\rm th}$ }}
\def\cth{{$c{\rm th}$ }}
\def\dth{{$d{\rm th}$ }}
\def\fth{{$f{\rm th}$ }}
\def\gth{{$g{\rm th}$ }}
\def\hth{{$h{\rm th}$ }}
\def\ith{{$i{\rm th}$ }}
\def\jth{{$j{\rm th}$ }}
\def\kth{{$k{\rm th}$ }}
\def\lth{{$l{\rm th}$ }}
\def\mth{{$m{\rm th}$ }}
\def\nth{{$n{\rm th}$ }}
\def\pth{{$p{\rm th}$ }}
\def\rth{{$r{\rm th}$ }}
\def\sth{{$s{\rm th}$ }}
\def\tth{{$t{\rm th}$ }}
\def\uth{{$u{\rm th}$ }}
\def\vth{{$v{\rm th}$ }}
\def\wth{{$w{\rm th}$ }}

\def\calA{{\mathcal A}}
\def\calB{{\mathcal B}}
\def\calC{{\mathcal C}}
\def\calD{{\mathcal D}}
\def\calE{{\mathcal E}}
\def\F{{\mathcal F}}
\def\G{{\mathcal G}}
\def\H{{\mathcal H}}
\def\calH{{\mathcal H}}
\def\IG{{\mathcal IG}}
\def\cI{{\mathcal I}}
\def\cP{{\mathcal P}}
\def\calJ{{\mathcal J}}
\def\calK{{\mathcal K}}
\def\calL{{\mathcal L}}
\def\calM{{\mathcal M}}
\def\calN{{\mathcal N}}
\def\N{{\calN}}
\def\calP{{\mathcal{P}}}
\def\calQ{{\mathcal{Q}}}
\def\calO{{\mathcal{O}}}
\def\calo{{\mathcal{o}}}
\def\calS{{\mathcal{S}}}
\def\calI{{\mathcal{I}}}
\def\calR{{\mathcal R}}
\def\calT{{\mathcal T}}
\def\calU{{\mathcal U}}
\def\calV{{\mathcal V}}
\def\calW{{\mathcal W}}
\def\calX{{\mathcal X}}
\def\calY{{\mathcal Y}}
\def\calZ{{\mathcal Z}}
\def\W{{\mathcal W}}
\def\U{{\Upsilon}}
\def\f{\frac}
\def\bX{{\bar X}}
\def\s{{\sigma}}
\def\p{{\partial}}
\def\p{{d}}
\def\v{{\varepsilon}}
\def\dis{{\displaystyle}}
\def\hash{{\#}}
\def\bar#1{{\overline{#1}}}
\def\hat#1{{\widehat{#1}}}
\def\barY{{\overline{Y}}}
\def\barX{{\overline{X}}}
\def\barZ{{\overline{Z}}}
\def\barx{{\overline{x}}}
\def\bary{{\overline{y}}}
\def\barz{{\overline{z}}}
\def\T{{ \mathrm{\scriptscriptstyle T} }}
\def\mT{{ -\mathrm{\scriptscriptstyle T} }}
\newcommand{\indep}{\perp\!\!\!\perp}
\newcommand{\nindep}{\perp\!\!\!\perp\!\!\!\!\!\!/\;\;}

\newcommand{\med}{{\rm med}}
\newcommand{\card}{{\rm card}}
\newcommand{\Splus}{\textsc{S-Plus}\ }


\def\rmi{{\rm i}}
\def\pr{{\rm Pr}}
\def\Pr{\pr}
\def\E{{\rm E}}
\def\Ev{{\rm Ev}}
\def\var{{\rm var}}
\def\cov{{\rm cov}}  
\def\corr{{\rm corr}}  
\def\Estar{{\rm E^*\thinspace}}
\def\varstar{{\rm var^*\thinspace}}
\def\covstar{{\rm cov^*\thinspace}}  
\def\median{{\rm median}}  
\def\cum{{\rm cum}}
\def\IMSE{{\rm IMSE}}  
\def\AIC{{\rm AIC}}  
\def\BIC{{\rm BIC}}  
\def\NIC{{\rm NIC}}  
\def\AICc{{\rm AIC_c}}  
\def\TIC{{\rm TIC}}  
\def\CLIC{{\rm CLIC}}  
\def\CV{{\rm CV}}  
\def\GCV{{\rm GCV}}  
\def\IQR{{\rm IQR}}  
\def\MAD{{\rm MAD}}  
\def\Lik{{L}}
\def\logL{{\ell}}
\def\RLik{{RL}}
\def\logLp{{\ell_{\rm p}}}

\def\Dto{{\ {\buildrel D\over \longrightarrow}\ }}
\def\Pto{{\ {\buildrel P\over \longrightarrow}\ }}
\def\rto{{\ {\buildrel r\over \longrightarrow}\ }}
\def\2to{{\ {\buildrel 2\over \longrightarrow}\ }}
\def\Wto{{\ {\buildrel W\over \longrightarrow}\ }}
\def\psto{{\ {\buildrel \rm{p.s.}\over \longrightarrow}\ }}
\def\asto{{\ {\buildrel \rm{a.s.}\over \longrightarrow}\ }}
\def\vto{{\ {\buildrel \rm{v}\over \longrightarrow}\ }}
\def\iid{{\ {\buildrel \rm{iid}\over \sim}\ }}
\def\ind{{\ {\buildrel \rm{ind}\over \sim}\ }}
\def\dotsim{{\ {\buildrel \cdot\over \sim}\ }}
\def\Deq{{\ {\buildrel D\over =}\ }}
\def\Eoneton{{$E_1,\ldots,E_n$}}
\def\I1ton{{$I_1,\ldots,I_n$}}
\def\X1ton{{$X_1,\ldots,X_n$}}
\def\Y1ton{{$Y_1,\ldots,Y_n$}}
\def\Z1ton{{$Z_1,\ldots,Z_n$}}
\def\R1ton{{$R_1,\ldots,R_n$}}
\def\e1ton{{$e_1,\ldots,e_n$}}
\def\t1ton{{$t_1,\ldots,t_n$}}
\def\x1ton{{$x_1,\ldots,x_n$}}
\def\y1ton{{$y_1,\ldots,y_n$}}
\def\z1ton{{$z_1,\ldots,z_n$}}


\def\AAP{\emph{Advances in Applied Probability}}
\def\AnnProbab{\emph{Annals of Probability}}
\def\AISM{\emph{Annals of the Institute of Statistical Mathematics}}
\def\AMS{\emph{Annals of Mathematical Statistics}}
\def\ApplStatist{\emph{Applied Statistics}}
\def\AS{\emph{Annals of Statistics}}
\def\B{\emph{Bernoulli}}
\def\Bka{\emph{Biometrika}}
\def\CJS{\emph{Canadian Journal of Statistics}}
\def\CSDA{\emph{Computational Statistics \& Data Analysis}}
\def\Eca{\emph{Econometrica}}
\def\ISI{\emph{International Statistical Review}}
\def\ISR{\emph{International Statistical Review}}
\def\JAMA{\emph{Journal of the American Medical Association}}
\def\JAP{\emph{Journal of Applied Probability}}
\def\JASA{\emph{Journal of the American Statistical Association}}
\def\JBES{\emph{Journal of Business \& Economic Statistics}}
\def\JCGS{\emph{Journal of Computational and Graphical Statistics}}
\def\JSCS{\emph{Journal of Statistical Computation and Simulation}}
\def\JRSSA{\emph{Journal of the Royal Statistical Society series A}}
\def\JRSSB{\emph{Journal of the Royal Statistical Society series B}}
\def\JRSSD{\emph{The Statistician}}
\def\SJS{\emph{Scandinavian Journal of Statistics}}
\def\SPA{\emph{Stochastic Processes and their Applications}}
\def\SM{\emph{Survey Methodology}}
\def\SS{\emph{Statistical Science}}
\def\TAS{\emph{The American Statistician}}
\def\Tech{\emph{Technometrics}}
\def\ZW{\emph{Z. Wahrsch. v. geb.}}
\def\JC{\emph{Journal of Classification}}
\def\PNAS{\emph{Proceedings of the National Academy of Science}}


\def\np{{\newpage}}
\def\gap{{\vskip 0.3in}}
\def\etal{{\sl et al.}} 
\def\pound{{\it \$}}
\def\apriori{{\sl a priori }}
\def\examfoot{{\vfill\hfill{\bf Turn over}\eject}}
\def\newtransparency{{\newpage}}
\def\Section{{Section~}}
\def\Sections{{Sections~}}
\def\Pvalue{{p-value}}
\def\Pvalues{{p-values}}
\def\pvalue{{p-value}}
\def\pvalues{{p-values}}

\def\report#1{{\newpage\centerline{\bf #1}\smallskip}}


\def\redbf#1{{\color{red}\bf #1}}%
\def\Defn{\redbf{Definition:\ }}%
\def\Thm#1{\redbf{Theorem #1:\ }}%
\def\Ex#1{\redbf{Example #1:\ }}%
\def\Note{\redbf{Note:\ }}%
\def\Exercise#1{\redbf{Exercise #1:\ }}%
\def\Ill{\redbf{Illustration:\ }}%
\def\calP{{\mathcal{P}}}
\def\calS{{\mathcal{S}}}
\def\calI{{\mathcal{I}}}
\def\eme{{\`eme}}
\def\ere{{\`ere}}
\def\picbox#1#2#3{{\centerline{\psfig{figure=#1,height=#2pc,angle=#3}}}}%
\def\pic#1#2#3{{\centerline{\includegraphics[height=#2pc,angle=#3]{#1}}}}%
\def\tt#1{{\texttt{#1}}}
\def\Der#1{\textbf{D#1}}

\def\signedAD{
\vspace{-1.5cm}\hspace{8.5cm}
\includegraphics[width=6cm]{/Users/davison/Dropbox/Originals/TeXACD/ADsignature}}

\def\signedA{
\vspace{-1.8cm}\hspace{8.5cm}
\includegraphics[width=4cm]{/Users/davison/Dropbox/Originals/TeXACD/Asignature}}

\def\o{{\mathrm{o}}}
%

%
%

\section{Introduction}

In a series of papers starting in the late 1980s, D.~A.~S.~Fraser, N.~Reid and coworkers developed  the tangent exponential model for higher-order likelihood inference. This chapter aims to explain the  motivation and justification for this model and to describe how it is used to compute accurate approximations.  The literature on this is not entirely transparent, as the argument evolved over numerous articles \citep{Fraser:1988,Fraser:1990,Fraser:1991,Fraser:2004,Fraser.Reid:1988,Fraser.Reid:1993,Fraser.Reid:1995,Fraser.Reid:2001,Cakmak.Fraser.Reid:1994,Fraser.Reid.Wu:1999}. We give a heuristic account of this construction, for the most part skating over the technical details, and provide an annotated bibliography as a road map through the literature.  The high accuracy of the resulting approximations has been verified empirically both in numerous articles and in books such as  \citet{Brazzale.Davison.Reid:2007}, Chapter~8 of which overlaps with the account here.

Approximations based on the tangent exponential model build on the theory of conditional and marginal inference, and on Laplace and saddlepoint approximations. Chapter~12 of \citet{Davison:2003} defines some basic notions and derives some of the results presented here.  Section~11.3.1 of that book contains an account of the Laplace method for integrals and related approximations for cumulative distribution functions.  Fuller accounts may be found in \citet{Barndorff-Nielsen.Cox:1989,Barndorff-Nielsen.Cox:1994}, \citet{McCullagh:1987}  and \citet{Severini:2000}.   \citet{Jensen:1995} and \citet{Butler:2007} provide comprehensive accounts of saddlepoint approximations and their many applications in analysis, probability, and statistics, and a helpful derivation is given in \citet{Kolassa:2006}. 

To fix notation and provide some building blocks that will be useful later, we start by summarising first-order inferential approximations related to the normal distribution and the central limit theorem. We then outline how the principles of sufficiency and ancillarity lead to consideration of significance functions, which are central in this and other frameworks for statistical inference, before discussing how the likelihood function can be viewed as a pivot. This leads to two density approximations, the $p^*$ approximation of \citet{BN1983} and the tangent exponential model density approximation of \citet{Fraser:1988}. Higher-order approximations to significance functions are then developed; first for linear exponential family models, where the link to saddlepoint approximations is most explicit, and then for general models, where the tangent exponential model plays a crucial role. Some extensions and generalisations are sketched in the concluding section.

\section{Likelihood and significance functions}

\subsection{Background}\label{background.subsect}

We consider a vector $Y=(Y_1,\ldots, Y_n)^\T$ of continuous responses and a statistical model for $Y$ with joint density function $f(y;\theta)$ that depends on a parameter $\theta\in\Theta\subset\mathbb{R}^p$. Vectors throughout are column vectors, and the superscripts $^\T$ and  $^\o$ denote transpose and a quantity evaluated at the observed data, so $y$ denotes a generic response vector and $y^\o$ its observed value, and $\hat\theta$ a maximum likelihood estimator and $\hat\theta^\o$ the maximum likelihood estimate computed from~$y^\o$. Continuity of the response distribution plays a crucial role in the general development; some comments on extending the tangent exponential model for discrete responses are given in Section~\ref{V.sect}.

The likelihood function is proportional to the density of the observed data regarded as a function of the unknown parameter $\theta$, i.e., 
$$
L(\theta) = L(\theta; y^\o) \propto f(y^\o;\theta),
$$
and measures the relative plausibility of different values of $\theta$ as explanations of $y^\o$.  Weighted by prior information, the likelihood is a main ingredient in Bayesian approaches to inference, and, unweighted, it is key to the ``pure likelihood'' approach \citep{Edwards:1972,Royall:1997}.  A central issue in using the likelihood function for inference is the distribution of $L(\theta;y)$ in repeated sampling under $f(y;\theta)$. This is needed in order to \emph{calibrate}\ inferences based on the likelihood function, i.e., to ensure that their stated properties are correct  under repeated sampling from the model.  Calibration is essential to give inferences objective validity, ideally while respecting basic principles of inference, such as conditionality and sufficiency.  One important approach to calibration is through the notion of a significance function, to be developed in Section~\ref{exact.subsect}.

We denote the log likelihood by $\ell(\theta) = \log L(\theta)$, and derivatives by subscripts, such as $\ell_\theta(\theta) = \partial\ell(\theta)/\partial\theta$ and $\ell_{\theta\theta}(\theta) = \partial^2\ell(\theta)/\partial\theta\partial\theta^\T$, which are respectively a column vector and a matrix. The observed information function is $\jmath(\theta) = -\ell_{\theta\theta}(\theta)$.  The classical asymptotic theory for likelihood-based inference is derived under the following smoothness conditions on the model \cite[Section~4.4.2]{Davison:2003}:
\begin{description}
\item{(i)} the true value of $\theta$ is interior to the parameter space $\Theta$;
\item{(ii)} the densities $\{f(y;\theta): \theta\in\Theta\}$ are distinct and have common support;
\item{(iii)} there is a neighbourhood ${\cal N}$ of the true value of $\theta$ within which the first three derivatives of $\ell(\theta)$ with respect to $\theta$ exist, and for $j,k,l = 1, \dots, p$, $n^{-1}\E\{|\ell_{\theta_j\theta_k\theta_l}(\theta)|\}$ is uniformly bounded for $\theta\in{\cal N}$;
\item{(iv)} the expected Fisher information matrix $I(\theta) = \E \{\jmath(\theta)\}$ is finite and positive definite, and $I(\theta) = \E\{\ell_{\theta}(\theta)\ell_{\theta}^{\T}(\theta)\}$. 
\end{description} 
Chapter~16 of \citet{vandervaart} gives weaker conditions for the limiting distributional results now described.

When $\theta$ is a scalar, this classical theory provides three basic distributional approximations for inference:
\begin{eqnarray} 
s = s{(\theta; y)} &=& 
\ell_\theta(\theta)\jmath^{-1/2}(\hat\theta) \dotsim N(0,1), \label{eq.rao}\\
t = t{(\theta; y)} &=& (\hat\theta - \theta)\jmath^{1/2}(\hat\theta) \dotsim N(0,1), \label{eq.wald} \\
r = r{(\theta; y)} &=& \text{sign}(\hat\theta - \theta)[2\{\ell(\hat\theta) - \ell(\theta)\}]^{1/2} \dotsim N(0,1), \label{eq.lrt}
\end{eqnarray}
where  the maximum likelihood estimator, $\hat\theta$, is assumed to satisfy $\ell_\theta(\hat\theta) =0$. 
The approximations~\eqref{eq.rao}--\eqref{eq.lrt} are derived from the central limit theorem for $\ell_\theta(\theta)$, which is $O_p(n^{1/2})$ in independent sampling, as  $\E_\theta\{\ell_\theta(\theta)\} = 0$ and $\var_\theta\{\ell_\theta(\theta)\} = O(n)$ under conditions (i)--(iv). Their derivations also rely on the consistency of $\hat\theta$  for $\theta$, and the convergence of the observed Fisher information to its expectation.

Each of $s$, $t$ and $r$ depends on both $\theta$ and the data $y$. The distributional approximations in~\eqref{eq.rao}--\eqref{eq.lrt} are all with reference to the distribution of $y$ under the model $f(y;\theta)$, and $s(\theta;y)$, $t(\theta;y)$ and $r(\theta;y)$ are approximate pivots, i.e., functions of the data and a parameter whose distribution is known, at least approximately. Thus if $y$ is fixed at $y^\o$, then $s(\theta;y^\o)$, $t(\theta;y^\o)$ and $r(\theta;y^\o)$ can be used to obtain the significance functions $\Phi(s)$, $\Phi(q)$, or $\Phi(r)$, where $\Phi(\cdot)$ is the distribution function for a standard normal random variable.  

Versions of~\eqref{eq.rao},~\eqref{eq.wald} and~\eqref{eq.lrt} are also available for vector-valued $\theta$.  We write $\theta = (\psi, \lambda)$, where $\psi$ is a scalar parameter of interest and $\lambda$ is a nuisance parameter. The profile log likelihood $\ell_{\text p}(\psi) = \ell(\psi, \hat\lambda_\psi)$ can be used to define analogous pivotal quantities,
\begin{eqnarray}
\label{eq.prof}	
	s &=& \ell_{{\text p}}'(\psi)\jmath_{\text p}^{-1/2}(\hat\psi) \dotsim N(0,1), \nonumber \\
	t &=& (\hat\psi - \psi)\jmath_{\text p}^{1/2}(\hat\psi) \dotsim N(0,1), \\
	r &=& \text{sign}(\hat\psi - \psi)[2\{\ell_{\text p}(\hat\psi) - \ell_{\text p}(\psi)\}]^{1/2} \dotsim N(0,1), \nonumber
\end{eqnarray}
where the prime denotes differentiation with respect to $\psi$, $\hat\lambda_\psi$ is the constrained maximum likelihood estimator for $\lambda$ for $\psi$ fixed, and 
$$
\jmath_{\rm p}(\psi)=\jmath_{\psi\psi}(\hat\theta_\psi) - \jmath_{\psi\lambda}(\hat\theta_\psi)\jmath_{\lambda\lambda}^{-1}(\hat\theta_\psi)\jmath_{\lambda\psi}(\hat\theta_\psi).
$$ 
We use the shorthand $\hat\theta_\psi = (\psi,\hat\lambda_\psi^{\T})^{\T}$, and partition the observed information matrix as
$$
\jmath(\theta)= \begin{pmatrix}\jmath_{\psi\psi}(\theta)&\jmath_{\psi\lambda}(\theta)\\ \jmath_{\lambda\psi}(\theta)&\jmath_{\lambda\lambda}(\theta)\end{pmatrix}, 
$$
with $\jmath_{\psi\lambda}(\theta) = -\partial^2\logL(\theta)/\partial\psi\partial\lambda^\T$, and so forth. 

The normal approximations to the distributions of $s$, $t$ and $r$ are equivalent to first order, but that for $r$ respects the asymmetry of the log likelihood about its maximum, whereas that for $t$ does not, which suggests that for complex problems these approximations may be inadequate. Their generally poor performance in models with many nuisance parameters has been borne out in empirical work.  

The approximations~\eqref{eq.rao}--\eqref{eq.prof} are usually derived from Taylor series expansion of the score equation $\ell_\theta(\hat\theta) = 0$ that defines the maximum likelihood estimate. A more elegant, but more difficult, approach shows that the log likelihood, as a function of $\theta$, converges to the log likelihood for a normal distribution. See, for example \citet{Fraser.McDunnough:1984} for a Taylor-series type approach to this and \citet{Lecam} or \citet[Ch.~6]{vandervaart} for a related approach that LeCam called ``local asymptotic normality". We will return to this in describing the $p^*$ approximation in Section~\ref{pstar.sect}.

In some settings we may be able to construct exact significance functions, using principles of sufficiency and ancillarity, which we now describe.

\subsection{Significance functions}\label{exact.subsect}

There are two ideal settings for exactly calibrating inference on a scalar parameter $\theta$.  In the first,  $Y$ can be reduced to a scalar minimal sufficient statistic $S$, giving the \emph{significance function}, sometimes also known as a \emph{\pvalue\ function},
\begin{equation}\label{sigfun1.eq}
p^\o(\theta) = \pr(S\leq s^\o;\theta), 
\end{equation}
which without loss of generality we suppose to be decreasing in $\theta$.  The quantity $P^\o(\theta)$ obtained by treating $s^\o$ as random has a uniform distribution under sampling from the model $f(y;\theta)$, and this allows the calibration of inferences described in Section~\ref{background.subsect}.  For example,  the limits of a $(1-2\alpha)$ confidence interval  $(\theta_\alpha,\theta_{1-\alpha})$ for $\theta$ are the solutions of equations $p^\o(\theta_\alpha) = 1-\alpha$ and $p^\o(\theta_{1-\alpha}) = \alpha$.  Likewise, the evidence against the hypothesis that $\theta=\theta_0$ with alternative $\theta\neq\theta_0$ may be summarised by the \Pvalue\ $ 2\min\{p^\o(\theta_0),1-p^\o(\theta_0)\}$.  The exact uniform distribution of $P^\o(\theta)$  under repeated sampling from $f(y;\theta)$ implies that both of these inferential summaries are perfectly calibrated, in the sense that they have exactly their stated properties: the confidence interval contains the true $\theta$ with probability $1-2\alpha$ and the \Pvalue\ is uniformly distributed.

In the second ideal setting there is a known transformation from $Y$ to a minimal sufficient statistic $(S,A)$, where $S$ is scalar and $A$ is \textit{ancillary},\ i.e., its distribution does not depend on $\theta$.  The role of $A$ can be understood by noting that the density $f(y;\theta)$ factorises as $f(s\mid a;\theta) f(a)$.  We can therefore envisage the data as arising first by generating $A$ and then generating $S$ conditional on $A$.  But as the first step does not depend on $\theta$, the relevant subset of the sample space for inference about $\theta$ fixes the observed value $a^\o$ of $A$ \citep{Cox:1958}.  Thus we should base inference for $\theta$ on the significance function
\begin{equation}\label{sigfun2.eq}
p^\o(\theta; a^\o) = \pr(S\leq s^\o\mid A=a^\o;\theta),
\end{equation}
which provides perfectly calibrated inferences, conditionally on the observed value of the ancillary statistic $A$.  Now the limits of an exact $(1-2\alpha)$ confidence interval for $\theta$, the solutions of equations $p^\o(\theta; a^\o)=\alpha$ and $p^\o(\theta;a^\o) = 1-\alpha$, depend on $a^\o$, thus emphasising how conditioning on the ancillary statistic $A$ affects the precision of inferences.

Expressions~\eqref{sigfun1.eq}  and~\eqref{sigfun2.eq} are functions of $\theta$, evaluated only at the observed data. 

\subsubsection*{Example 1}

Suppose that $Y_1/\theta$ and $Y_2\theta$ are independent gamma variables with unit scale and shape parameter $n$; their joint density function is
$$
f(y_1,y_2;\theta) = { (y_1y_2)^{n-1}\over \{\Gamma(n)\}^2} \exp\left( -y_1/\theta - y_2\theta\right), \quad y_1, y_2>0, \theta>0, 
$$
and the minimal sufficient statistic is $(Y_1,Y_2)$.  We set $S=(Y_1/Y_2)^{1/2}$ and $A=(Y_1Y_2)^{1/2}$, which is ancillary, and note  that since $Y_1=AS$ and $Y_2=A/S$, we have $y^\T=(y_1,y_2) = (as,a/s)$, so $|\partial(y_1,y_2)/\partial(a,s)|=2a/s$ and 
$$
f(s\mid a;\theta) = {1\over s I(a)}\exp\left\{-a( s/\theta + \theta/s)\right\}, \quad a,s>0, \theta>0, 
$$
where $I(a)=\int_{-\infty}^\infty \exp( - 2a\cosh u)\, \D{u}$ is a normalising constant.  The significance function~\eqref{sigfun2.eq} is readily obtained by numerical integration of $f(s\mid a^\o;\theta)$ over the interval $(0,s^\o)$.  

The log likelihood for this model can be written as 
$$
\logL(\theta; s, a) \equiv - a ( s/\theta + \theta/s), \quad \theta>0, 
$$
so the maximum likelihood estimator is $\hat\theta=s$ and the observed information is $\jmath(\hat\theta) = a/s^2$. Larger values of the ancillary $a$ yield more precise inferences, because the standard error for $\hat\theta$, $\jmath^{-1/2}(\hat\theta) = s/a^{1/2}$, diminishes as $a$ increases.  

Figure~\ref{fig0} shows the significance and log likelihood  functions when $s^\o=1.6$ and $a^\o=3$, 6.  They are more concentrated when $a^\o=6$, resulting in shorter confidence intervals. Both log likelihood  functions show clear asymmetry, suggesting that normal approximation based on $t$ is badly calibrated and would provide poor inferences; indeed, the symmetric approximation~\eqref{eq.wald} could produce negative confidence limits. 

\hfill \endex

\begin{figure}[t]
\begin{center}
\centerline{\includegraphics[width=\textwidth]{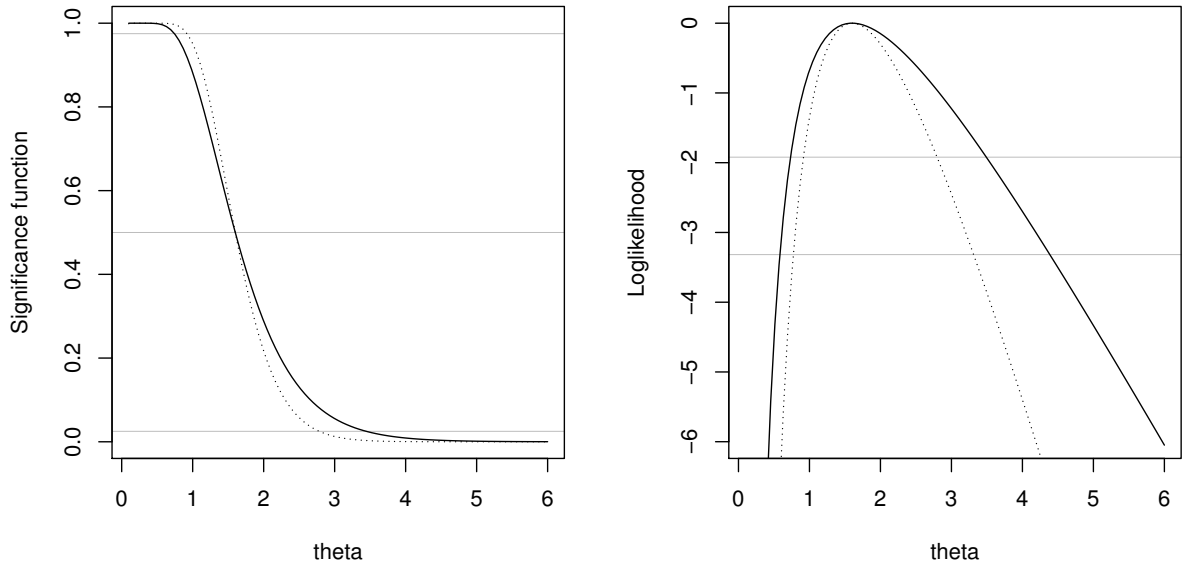}}
\caption{Inference for gamma example.  Significance functions $\pr(s\leq s^\o\mid a;\theta)$ (left) and log likelihoods (right) for $a=3$ (solid) and $a=6$ (dots). The horizontal lines in the left-hand panel are at 0.025, 0.5 and 0.975; the intersections of the highest and lowest of these with the significance functions give the limits of exact 95\% confidence intervals.  The horizontal lines in the right-hand panel are at $-1.92$ and $-3.32$, and their intersections with the log likelihoods show the limits of 95\% and 99\% confidence intervals based on the approximation~\eqref{eq.lrt}.} 
\label{fig0}
\end{center}
\end{figure}

Though rarely met in practice, the above settings provide blueprints for more complex situations, in which several complications may arise:
\bi
\item $\theta$ consists of a parameter $\psi$ of interest, often scalar, and a vector $\lambda$ of nuisance parameters.  Even if there is a direct analogue of $S$, the significance functions~\eqref{sigfun1.eq} and~\eqref{sigfun2.eq} will typically depend on the unknown $\lambda$;
\item the reduction to a minimal sufficient statistic of dimension $p$ applies only in {linear} exponential family models.  No mapping $y \mapsto (s,a)$ can be found in general, so exact inferences are mostly unavailable; and 
\item the interest parameter $\psi$ may be a vector. We shall not consider this situation here, but Section~\ref{ext.sect} has pointers to related approaches.
\ei

Significance functions were emphasized as a primary tool for inference in \cite{Fraser:1991}. \cite{Fraser:2019} describes the many summaries that are then directly available: confidence limits and \Pvalues,\ as described above, as well as aspects of fixed-level testing. Power, for example, is reflected in the ``steepness'' of the function. There is a close connection between significance functions and confidence distributions \citep{Cox:1958, Efron:1993, Xie.Singh:2013}.

\subsection{Likelihood as pivot}\label{pstar.sect}

A fruitful approach to improved approximations is to use the log likelihood more directly---\citet{Hinkley:1980} referred to the likelihood function itself as a pivot.  One reason this provides more precise calibration of our inferences is that it is closely related to a saddlepoint approximation \citep{Daniels:1954}, which can be uncannily accurate.  The saddlepoint approximation can be applied to the density of the maximum likelihood estimator, leading to the  $p^*$ \textit{approximation},  also called ``Barndorff-Nielsen's formula'' \citep{BN1980}: 
\begin{equation}\label{eq.pstar}
p^*(\hat\theta \mid a; \theta) = c |\jmath(\hat\theta)|^{1/2}\exp\{\ell(\theta)-\ell(\hat\theta)\},	
\end{equation}
where $\ell(\theta) = \ell(\theta;\hat\theta,a)$, with $(\hat\theta,a)$ a transformation of the sample $y=(y_1, \dots, y_n)$ such that $a$ is ancillary.  Pivoting this approximate density~\eqref{eq.pstar}, which is supported on $\mathbb{R}^p$, provides confidence intervals or regions for $\theta$, and since $a$ is ancillary no information about the parameter is lost by conditioning. However to compute the approximation requires calculation of the transformation from $y$ to $(\hat\theta,a)$, which can be very difficult.  Moreover, computation of a significance function requires only a good approximation at the observed data point, whereas using~\eqref{eq.pstar} would require high accuracy for all $(\hat\theta,a)$. 

An approach intermediate between using the limiting normal form for the log likelihood and using~\eqref{eq.pstar} was introduced in \citet{Fraser:1988}.  His \textit{tangent exponential model}\  approximation to the density $f(y;\theta)$, defined for $y\in \mathbb{R}^n$ and $\theta\in \mathbb{R}^p$, is a model on $\mathbb{R}^p$ that implements conditioning on an approximately ancillary statistic $a$.  Its expression is
\begin{equation}\label{eq.tem}
f_{\rm TEM}(s \mid a;\theta) = \exp[s^\T\varphi(\theta) + \ell\{\theta(\varphi); y^\o\}]h(s).
\end{equation}
This has the structure of a linear exponential family model for a constructed sufficient statistic $s\in\mathbb{R}^p$ 
and constructed canonical parameter $\varphi(\theta)\in\mathbb{R}^p$,  with $-\ell(\theta;y^\o)$ playing the role of the cumulant generator. 
   If the underlying density is in the exponential family, then $\varphi(\theta)$ is simply the canonical parameter. In more general models the canonical parameter $\varphi(\theta)$ may depend on the data $y^\o$, and is constructed using principles of approximate ancillarity, as described in Section~\ref{basic.sect}.

In the next section we show how this tangent exponential model is built on the exact significance functions of Section~\ref{exact.subsect}, but incorporates aspects of direct approximation of the log likelihood.  

\section{Approximate conditional inference}\label{approx.sect}

\subsection{Ancillary and sufficient directions}
 
Below we describe a general approach to approximate but accurate inference when the dimension $p$ of the parameter vector $\theta$ is less than the dimension $n$ of the data $y$.  We assume that the components of $y$ are independent and that the regularity conditions outlined in Section~\ref{background.subsect} hold. The tangent exponential model is often said to have ``asymptotic properties'', which essentially is shorthand for the assumption that the expansions in Section~\ref{derivation} have the behaviour in $n$ indicated there, i.e., the log-density is differentiable in both $\theta$ and $y$ up to quartic terms, and the standardized derivatives decrease in powers of $n^{-1/2}$.

For a heuristic development we suppose initially that there is a smooth bijection between $y$ and $(s,a)$, where $s$ and $a$ have respective dimensions $p$ and $n-p$, and $a$ is ancillary.  The conditionality principle implies that inference should be based on the conditional density $f(s\mid a^\o;\theta)$,
so  the reference set for frequentist inference is $\calA^\o=\{y\in\Real^n: a(y) = a^\o\}$, i.e.,  the $p$-dimensional manifold of the sample space on which the ancillary statistic equals its observed value $a^\o$.  
The bijection between $y$ and $(s,a)$ implies that $\calA^\o$ can be parametrised in terms of $s$, at which point its tangent plane $\calT_s$ is determined by the columns of the $n\times p$ matrix 
\begin{equation}\label{v0.eq}
{\partial y(s,a^\o)\over \partial s^\T}.
\end{equation}
In particular, the tangent plane $\calT^\o$ to $\calA^\o$ at $s^\o$ is determined by
\begin{equation}\label{v1.eq}
V = \left.{\partial y(s,a^\o)\over \partial s^\T}\right|_{s=s^\o}.
\end{equation}
The columns of $V$ have been called ancillary directions, because they are derived from the ancillary manifold $\calA^\o$ at $s=s^\o$, but one can argue that this is a misnomer: the $p$ columns of $V$ determine how $y$ changes in the direction of $s$ locally at $s^\o$, so they might better be called \textit{sufficient directions},\ and we shall use this term below.  The ancillary statistic itself varies locally at $a^\o$ in the $n-p$ directions orthogonal to the columns of $V$. 

Suppose for simplicity that $\theta$ and $S$ are scalar.  Then 
\begin{equation}\label{integral.eq}
\pr(S\leq s^\o\mid A=a^\o;\theta) \, \propto\,  {\int_{-\infty}^{s^\o} f(s,a^\o;\theta)\, \D{s}}.
\end{equation}
The log likelihood $\logL(\theta;s,a) = \log f(s,a;\theta)$ is a sum of $n$ contributions and therefore has order $n$.  A change of variables can be used to ensure that $s$ is $O_p(1)$ and then, for $s=s^\o +O_p(n^{-1/2})$, Taylor series expansion gives 
\begin{equation}\label{exp-family.eq}
\logL(\theta;s,a^\o) = \logL(\theta;s^\o,a^\o) + (s-s^\o) \left.{\partial \logL(\theta;s,a^\o) \over \partial s}\right|_{s=s^\o} + \cdots,
\end{equation}
where the first term on the right-hand side is of order $n$, the second is the product of a term of order $n^{-1/2}$ with one of order $n$ and is therefore of order $n^{1/2}$, and the remainder is $O(1)$. {Standard first-order results, such as that leading to a significance function from applying the asymptotic approximation~\eqref{eq.lrt} to the signed likelihood root
$$
r^\o(\theta) = \sign(\hat\theta^\o-\theta)\left[2\left\{\logL(\hat\theta^\o;s^\o,a^\o) - \logL(\theta;s^\o,a^\o)\right\}\right]^{1/2}, 
$$
 use only the first term on the right of~\eqref{exp-family.eq}, and have error of order $n^{-1/2}$ for one-sided confidence intervals.} We hope to reduce this error to order $n^{-1}$, so-called second-order inference, by including the second term.  Although $r(\theta)$ involves only the value of the log likelihood, an approximation based on both terms in~\eqref{exp-family.eq}  also requires the derivative of $\logL$ with respect to $s$, a so-called sample space derivative. Approximation \eqref{exp-family.eq} yields a version of the tangent exponential model (\ref{eq.tem}), here specialized to the case where we can identify $(s,a)$ directly.
 
    In Section~\ref{exp-family:sect}  we describe the building blocks for exponential family models, but we first return to the example in Section~\ref{exact.subsect}. 

\subsubsection*{Example 1 (ctd)}

We previously saw that $y^\T(s,a)=(y_1,y_2) = (as,a/s)$, yielding
$$
 {\partial y(s,a)\over \partial s}  
= \begin{pmatrix}  a\\ -a/s^2\end{pmatrix} , \quad V = \begin{pmatrix}  a^\o\\ -a^\o/s^{\o2}\end{pmatrix} , \quad 
{\partial \logL(\theta; s, a)\over \partial s} = -a(\theta^{-1} - \theta s^{-2}).
$$
The left-hand panel of Figure~\ref{fig1} shows $f(y_1,y_2;\theta)$, the conditional sample space $\calA^\o$ and its tangent space $\calT^\o$ for $a^\o=3$ and $s^\o=1.6$.  The right-hand panel shows the logarithm of the conditional density $f(s\mid a^\o;\theta)$ and its tangents at $s^\o$ for $\theta$ equal to $1$, to $\hat\theta^\o=1.6$ and to 2.2.  The filled circles show $f(s^\o\mid a^\o;\theta)$  for these values of $\theta$, i.e., the corresponding likelihood values.  Notice that $\theta = \varphi b + \{ \varphi^2 b^2 + (s^\o)^2\}^{1/2}$,  where $b=(s^\o)^2/(2a^o)$, can be expressed in terms of $\varphi(\theta)=\partial\logL(\theta; s^\o,a^\o)/\partial s$, thus parametrising the model in terms of the slope of the tangent to the log density at $y^\o$; in this model this is a data-dependent parametrisation, to which we return below.\endex

\begin{figure}[t]
\begin{center}
\centerline{\includegraphics[width=\textwidth]{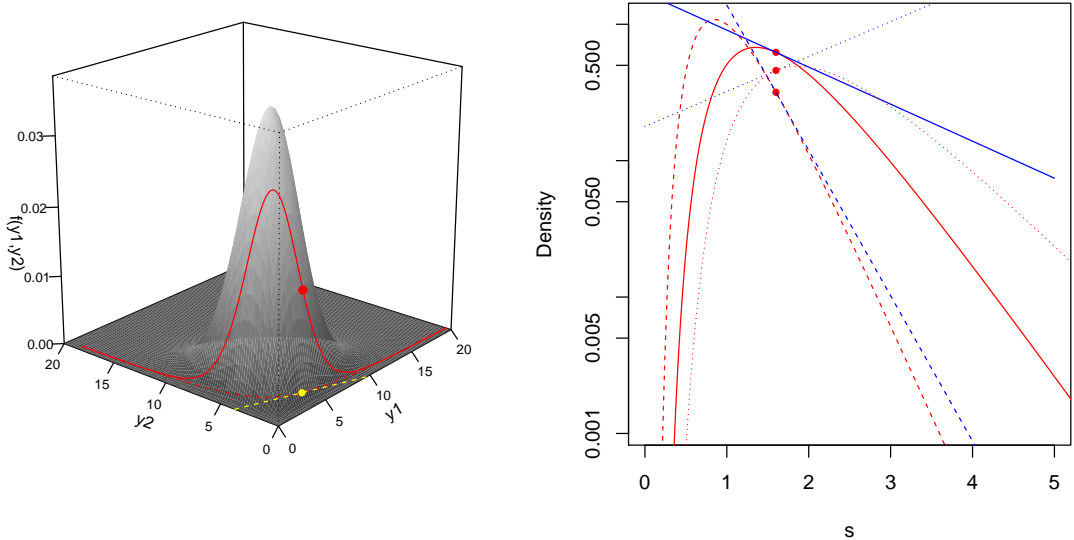}}
\caption{Conditional inference for gamma example.  Left: joint density $f(y;\theta)$ with $\theta=1$, $y^\o = (4.8, 1.875)$, giving $a^\o=3$ and $s^\o=1.6$. The yellow filled circle shows $y^\o$ and the red filled circle shows $f(y^\o; \theta)$. The conditional reference set $\calA^\o$ is shown by the dashed red line and the density $f\{y(s,a^\o);\theta\}$ on $\calA^\o$ is shown by the solid red line.    The tangent plane $\calT^\o$ is shown by the dashed yellow line.  Right: Log conditional density $f(s\mid a^\o;\theta)$ (red) and its tangents at $\log f(s^\o\mid a^\o;\theta)$ (blue) for $\theta=\hat\theta^\o$ (solid), $\theta=1$ (dashes) and $\theta=2.2$ (dots). The red filled circles are at $f(s^\o\mid a^\o;\theta)$.} 
\label{fig1}
\end{center}
\end{figure}

\subsection{Computation of $V$}\label{computation:sect}

At first sight the definition of $V$ at~\eqref{v1.eq} suggests that the mapping $y\mapsto(s,a)$ must be known.  In fact this is not the case, as we see if we write
\begin{equation}
\label{V.eq1}
V = \left.{\partial y\over\partial s^\T}\right|_{y=y^\o} 
= \left. {\partial y\over\partial \theta^\T}\right|_{y=y^\o}
\times\left.\left({\partial s\over\partial \theta^\T}\right)^{-1}\right|_{y=y^o}.  
\end{equation}
The second matrix on the right has dimension $p\times p$ and {is invertible}, so the column space of $V$ is also the column space of the first matrix on the right; both define the same space of sufficient directions, but $\partial y/\partial \theta^\T$ does not require $y$ to be expressed in terms of $(s,a)$.   Hence $V$ could be right-multiplied by any invertible $p\times p$ matrix of constants without 
changing the span of the sufficient directions.  Below we shall generally take $V$ to be $\partial y/\partial \theta^\T$, evaluated at $y=y^\o$ and $\theta=\hat\theta^\o$. 

Although it is not customary to express $y$ as a function of $\theta$, it is natural to do so when considering how a dataset would be simulated.  In a regression model, for example, we can write $\theta=(\beta,\s)$ and
\begin{equation}
\label{linear.eq}
y(\theta) = X\beta + \s\v,
\end{equation}
with the design matrix $X$ and error vector $\v$  fixed, giving 
\begin{equation}\label{V-linear-model.eq}
V = \left. \left( {\partial y\over\partial \beta^\T}, {\partial y\over\partial \sigma}\right)\right|_{y=y^\o}  = \left.(X, \v)\right|_{y=y^\o} = (X, (y^\o-X\hat\beta^\o)/\hat\s^\o),
\end{equation}
where $\hat\theta^\o=(\hat\beta^\o,\hat\s^\o)$ are maximum likelihood estimates computed from the data $y^\o$.  One way to think about~\eqref{linear.eq} is as a quantile function, or structural equation, whereby changes in $\theta$ are reflected in changes to $y$ for fixed $\v$. 

An alternative approach to deriving $V$ is to note that if the $y_j$ are independent and have distribution functions $F_j(\cdot;\theta)$, then  the pivotal equation $F(y_j;\theta)=u_j$ for some fixed $u_j$ implicitly defines how $y_j$ depends on $\theta$. Total differentiation of this equation with respect to $\theta^\T$ yields
$$
0= {\D{u_j}\over \D{\theta^\T}} = {\partial y_j\over \partial \theta^\T}{\partial F(y_j;\theta)\over \partial y_j} + {\partial F(y_j;\theta)\over \partial \theta^\T}, 
$$
which implies that
\begin{equation}\label{altV.eq}
{\partial y_j\over\partial  \theta^\T} = -\left\{ {\partial F_j(y_j;\theta)\over\partial y_j}\right\}^{-1}{\partial F_j(y_j;\theta)\over\partial \theta^\T}.
\end{equation}
In the case of a regression model we have
$$
F_j(y_j;\beta,\s) = F\{ (y_j-x_j^\T\beta)/\s\},
$$
where $x_j^\T$ is the $j$th row of $X$ and $\v_1,\ldots, \v_n\iid F$, which gives~\eqref{V-linear-model.eq} when evaluated at $y^\o$. 

In group transformation models there is no need to invoke the distribution function $F_j$, because one can write $y_j=g_j(\theta,\v_j)$ for a known function $g_j$ and an error $\v_j$ whose distribution does not depend on $\theta$. Then $V$ can be computed as $\partial y_j/\partial \theta$ for fixed $\v_j$, or equivalently
$$
{\partial y_j\over\partial  \theta^\T} = -\left\{ {\partial \v_j(y_j;\theta)\over\partial y_j}\right\}^{-1}{\partial \v_j(y_j;\theta)\over\partial \theta^\T}.
$$
This construction is used in the bivariate normal example of Section~\ref{V.sect}.  The derivation of the sufficient directions in \citet{Fraser.Reid:1995} and \citet{Fraser.Reid:2001} builds on the local location model of \citet{Fraser:1964}; see Section~\ref{hist.sect}.

The construction in terms of a structural equation such as~\eqref{linear.eq} does not apply to discrete models, which require a slightly different treatment.  
We discuss the construction of the sufficient directions $V$ further in Section~\ref{V.sect}.

In the next section we outline the accurate approximation of significance functions in continuous exponential families.

\section{Exponential family inferences}\label{exp-family:sect}

\subsection{One-parameter case}\label{exp-family-one.sect}

Consider a continuous one-parameter exponential family with a scalar parameter $\theta$, expressed as a tilted version of a baseline density $f_0(y)$,

\begin{equation}
\label{exp-family.eq0}
f(y;\theta) = f_0(y)\exp\left[n\left\{s(y)\theta - \kappa(\theta)\right\}\right], 
\end{equation}
with canonical statistic $ns(y)$, canonical parameter $\theta$ and cumulant generator $n\kappa(\theta)$.  The marginal density of the sufficient statistic $s$ is of the same form,
\begin{equation}
\label{exp-family.eq01}
f(s;\theta) = h_0(s)	\exp\left[n\left\{s\theta - \kappa(\theta)\right\}\right],
\end{equation}
where $h_0(s) = \int 1\{y\in\mathbb{R}^n : s(y) = s\}f_0(y)dy$.
Here $s$ is assumed to be an average of $n$ independent observations, so its variation around its mean $\kappa_\theta(\theta)$ is $O_p(n^{-1/2})$.  Writing densities explicitly in terms of the sample size $n$, as in~\eqref{exp-family.eq0} or~\eqref{exp-family.eq01}, is a technical device that helps in keeping track of powers of $n$  in theoretical development, but it does not change the quality of subsequent approximations and is unnecessary when applying them.

Although the integral defining $h_0$ is not usually available in closed form, the saddlepoint approximation can be used to approximate $f(s;\theta)$ very accurately. And as  $\hat\theta$ satisfies $s = \kappa_\theta(\hat\theta)$, this gives an approximation to the density of $\hat\theta$ \cite[e.g.][equation~(12.32)]{Davison:2003}:
\begin{equation}
\label{exp-family.eq1}
f(\hat\theta; \theta) = c\,\hat \jmath^{~1/2}\exp\left\{\logL(\theta; \hat\theta) - \logL(\hat\theta; \hat\theta)\right\}\left\{1 + O(n^{-1})\right\}, 
\end{equation}
where $\hat \jmath = n\kappa_{\theta\theta}(\hat\theta)$ 
is the observed information  evaluated at $\theta=\hat\theta$, and the normalising constant $c$ ensures that~\eqref{exp-family.eq1} has unit integral.  Our goal below is to use~\eqref{exp-family.eq1} to obtain a convenient and general expression for the significance function
$\pr(\hat\theta \leq \hat\theta^\o;\theta)$, where $\hat\theta^\o$ is the observed value of $\hat\theta$. 

We first make a monotone change of variable $\hat\theta \mapsto r(\theta)$, where 
$$
r(\theta) = \sign(\hat\theta-\theta)\left[2\left\{\logL(\hat\theta; \hat\theta)-\logL(\theta; \hat\theta)\right\}\right]^{1/2}
$$
The Jacobian of this transformation may be obtained from the derivative 
\begin{eqnarray*}
	r(\theta){\partial r(\theta)\over\partial \hat\theta } &=& 
\left.{\partial \logL(\theta; \hat\theta)\over\partial\theta}\right|_{\theta=\hat\theta} + 
\left.{\partial \logL(\theta; \hat\theta)\over\partial\hat\theta}\right|_{\theta=\hat\theta}
-
{\partial \logL(\theta; \hat\theta)\over\partial\hat\theta}, \\
&=&\logL_{;\hat\theta}(\hat\theta;\hat\theta) - \logL_{;\hat\theta}(\theta;\hat\theta),
\end{eqnarray*}
where here and below the appearance of a variable after a subscripted semi-colon indicates a sample-space derivative with respect to that variable. Differentiation with respect to $\hat\theta$ is
necessitated by the change of variable $\hat\theta\mapsto r(\theta)$. For the exponential model we have
\begin{eqnarray*}
r(\theta){\partial r(\theta)\over\partial \hat\theta } &=& n\kappa_{\theta\theta}(\hat\theta)(\hat\theta - \theta) = \hat\jmath\, (\hat\theta-\theta),
\end{eqnarray*}
and~\eqref{exp-family.eq1} becomes 
\begin{equation}
\label{exp-family.eq2}
f\{r(\theta);\theta\} = c  {r(\theta)\over q(\theta)}  \exp\left\{ - r(\theta)^2/2\right\}, 
\end{equation}
where the Wald statistic for $\theta$,
$$
q(\theta) = \hat\jmath^{~1/2}(\hat\theta - \theta)= \hat\jmath^{-1/2}\left\{ \logL_{;\hat\theta}(\hat\theta;\hat\theta) - \logL_{;\hat\theta}(\theta;\hat\theta)\right\} , 
$$
has an asymptotic standard normal distribution; see~\eqref{eq.wald}, where $q$ is called $t$.

As $\theta\to\hat\theta$ we have $r(\theta), q(\theta)\to 0$, but it is possible to show that, using an abbreviated notation, 
$q = r + a_1 r^2/n^{1/2} + a_2 r^3/n +\cdots$, so there is no singularity in~\eqref{exp-family.eq2}. Taylor series expansion of the  logarithm yields 
\begin{equation}
\label{correction.eq}
{1\over r} \log\left({q\over r}\right) = a_1n^{-1/2} + (a_2-a_1^2/2)r n^{-1} + O(n^{-3/2}),
\end{equation} 
so the square of~\eqref{correction.eq}  is of order $n^{-1}$.   This is useful in the next step. 

It follows from~\eqref{correction.eq} that a second change of variable 
\begin{equation}\label{r-star.eq}
r(\theta)\, \mapsto\,  r^*(\theta)  = r(\theta) + {1\over r(\theta)}\log\left\{{q(\theta)\over r(\theta)}\right\}
\end{equation}
has Jacobian $1 + (a_2-a_1^2/2)/n + O(n^{-3/2}) = 1+O(n^{-1})$. That the coefficient of the $1/n$ term is constant in $r$ is important for renormalisation, discussed  in Section~\ref{renorm.sect}.  Although the transformation~\eqref{r-star.eq} may not be strictly monotonic over its entire range, it is monotone to the order considered here. After this transformation and a little more algebra,~\eqref{exp-family.eq2} becomes 
\begin{equation}
\label{exp-family.eq3}
f\{r^*(\theta);\theta\} = c   \exp\left\{ - r^*(\theta)^2/2\right\}\left\{1 + O(n^{-1})\right\}.
\end{equation}
We deduce that $c=(2\pi)^{-1/2}\{1+O(n^{-1})\}$, so the significance function may be written
\begin{eqnarray}
\nonumber
\pr(\hat\theta \leq \hat\theta^\o;\theta) &= &
\pr\left\{r^*(\theta)\leq r^{*\o}(\theta);\theta\right\} \\
&=&
\Phi\left\{r^{*\o}(\theta)\right\}\left\{1+O(n^{-1})\right\},\label{exp-family.eq4}
\end{eqnarray}
where $r^{*\o}(\theta)$ is the value of $r^*(\theta)$ actually observed, i.e., with $y$ and $\hat\theta$ replaced by $y^\o$ and $\hat\theta^\o$.  The discussion in Section~\ref{basic.sect} then allows inference on $\theta$ by treating $r^{*\o}(\theta)$ as a realisation of a standard normal variable.  An alternative to~\eqref{exp-family.eq4} with the same order of asymptotic error is the Lugannani--Rice (\citeyear{Lugannani.Rice:1980}) formula
\begin{equation}
\label{LR.eq}
\pr\left(\hat\theta \leq \hat\theta^\o;\theta\right) \doteq \Phi\left\{r^{\o}(\theta)\right\} + \left\{{1\over r^{\o}(\theta)} - {1\over q^{\o}(\theta)} \right\} \phi\left\{r^{\o}(\theta)\right\},
\end{equation}
where $\phi$ denotes the standard normal density function and $r^\o(\theta)$ and $q^\o(\theta)$ are the ingredients to $r^{*\o}(\psi)$.  Both~\eqref{exp-family.eq4} and~\eqref{LR.eq} are typically highly accurate, with neither systematically better than the other in applications, but they can become numerically unstable for $\theta$ {near $\hat\theta$ and dealing with this may require  some careful programming. }
In Section~\ref{renorm.sect} we show that when the density is renormalized to integrate to unity the relative error in \eqref{exp-family.eq4} and \eqref{LR.eq} becomes $O(n^{-3/2})$, as is that in \eqref{exp-family.eq3}.

Expression~\eqref{correction.eq} shows that the second term of $r^*(\theta)$ is an $O(n^{-1/2})$  correction to the $O(1)$ quantity $r(\theta)$; recall the second term on the right-hand side of~\eqref{exp-family.eq}.  

\subsection{Linear exponential family}\label{mult.sec}

The argument above generalises to a linear exponential family in which $\theta^\T=(\psi,\lambda^\T)$ consists of a scalar parameter $\psi$ of interest and a nuisance parameter $\lambda$ of dimension $p-1$. In this case,
$$
f(s;\theta) = h(s)\exp\left[n\left\{s^\T\theta - \kappa(\theta)\right\}\right],
$$
where $s(y)^\T = (s_1,s_2)$  is partitioned conformably with $\theta$.  In this model the conditional density of $s_1$ given $s_2$ does not depend on $\lambda$, and the ratio of the saddlepoint approximations to the densities of $(s_1,s_2)$ and of $s_2$ yields the approximation
\begin{equation}\label{expcond.eq}
f(s_1\mid s_2;\psi) \doteq c \left\{{|\jmath_{\lambda\lambda}(\hat\theta_\psi)|\over |\jmath(\hat\theta)|}\right\}^{1/2} \exp\left\{\logL(\hat\theta_\psi) - \logL(\hat\theta)\right\}, 
\end{equation}
where 
$\hat\theta_\psi= (\psi, \hat\lambda^\T_\psi)^\T$ denotes the maximum likelihood estimator for fixed $\psi$ and $\jmath_{\lambda\lambda}(\theta)$ is the sub-matrix of $\jmath(\theta)$ corresponding to $\lambda$.  A calculation similar to that leading to~\eqref{exp-family.eq3} establishes that apart from a  relative error of order $n^{-1}$,  the  resulting significance function for inference on $\psi$, $\pr(S_1\leq s_1^\o\mid S_2=s_2;\psi)$  is again of form~\eqref{exp-family.eq4}, now with
\begin{eqnarray}
\label{r-and-q.eq-r}
r(\psi) &=& \sign(\hat\psi-\psi)\left[2\left\{\logL(\hat\theta) - \logL(\hat\theta_\psi)\right\}\right]^{1/2}, \\
\label{r-and-q.eq-q}
 q(\psi) &=& (\hat\psi-\psi) \left\{{|\jmath(\hat\theta)|\over |\jmath_{\lambda\lambda}(\hat\theta_\psi)|}\right\}^{1/2}
\end{eqnarray}
evaluated at the observed data $y^\o$ and maximum likelihood estimate $\hat\theta^\o$.  Note that 
$$
q(\psi) = t(\psi)\rho(\psi,\hat\psi), \quad \rho(\psi,\hat\psi) = \left\{{|\jmath_{\lambda\lambda}(\hat\theta)|\over |\jmath_{\lambda\lambda}(\hat\theta_\psi)|}\right\}^{1/2},
$$
 with $t(\psi)$ the Wald statistic based on the profile log likelihood defined in \eqref{eq.prof}. For derivations of these results see \citet{Fraser.Reid:1993} or  \citet[\S12.3.3]{Davison:2003}, for example.

Although expression~\eqref{exp-family.eq4} was derived by approximating a conditional distribution, it is also an approximation to the marginal distribution of $r^*(\psi)$, because the normal distribution of $r^*(\psi)$ in~\eqref{exp-family.eq4} does not depend on the conditioning variable $S_2$.  

Moreover, if the parameter of interest is a linear function of the natural parameter $\theta$ of the exponential model, say $\psi=C_1^\T\theta$ for some known $p\times 1$ vector $C_1$, then we can set $\lambda=C_2^\T\theta$, so that $\varphi^\T  = (\psi,\lambda^\T) = \theta^\T (C_1, C_2) = \theta^\T C$, say, where the $p\times p$ matrix $C$ is invertible, express the exponential family in terms of 
$$
s_*(y) = C^{-1}s(y), \quad \varphi(\theta) = C^\T \theta, \quad \kappa_*(\varphi) = \kappa(C^{-\T}\varphi)=\kappa(\theta),
$$
and finally apply approximation~\eqref{exp-family.eq4} in this reparametrised model.

\subsection{General exponential family}\label{exp-family-general.sect}

We now extend the normal approximation to the distribution of $r^*$, developed in the previous section, to the general exponential family
\begin{equation}\label{general-exp-family.eq}
f(s;\theta) = h(s)\exp\left[n\left\{s^\T\varphi(\theta)-\kappa(\theta)\right\}\right],
\end{equation}
where $\varphi(\theta)$ may be a nonlinear function of $(\psi, \lambda)$. For this we shall need the analogues of $r(\psi)$ and  $q(\psi)$ of~\eqref{r-and-q.eq-r} and~\eqref{r-and-q.eq-q}.  The likelihood is invariant to reparametrisation, so $r(\psi)$ is unchanged, but as $\psi$ is no longer a component of the canonical parameter $\varphi$, we need a new form for $q(\psi)$ using a surrogate for $\psi$.
Taylor series expansion for small $\hat\theta-\hat\theta_\psi$ gives 
$$
\varphi(\hat\theta)-\varphi(\hat\theta_\psi) = {\partial\varphi(\hat\theta_\psi)\over \partial \theta^\T} (\hat\theta-\hat\theta_\psi) + \cdots, 
$$
so if the matrix on the right-hand side is invertible at $\hat\theta_\psi$, which is a condition for $\theta$ to be identifiable,  then 
\begin{equation}\label{inverse.eq}
\hat\theta-\hat\theta_\psi = \left\{{\partial\varphi(\hat\theta_\psi)\over \partial \theta^\T}\right\}^{-1} \left\{\varphi(\hat\theta)-\varphi(\hat\theta_\psi) \right\} + \cdots, 
\end{equation}
and as the partial derivatives satisfy
\begin{equation}\label{identity.eq}
I_p = {\partial\theta\over \partial\varphi^\T} {\partial \varphi\over \partial\theta^\T} = 
\begin{pmatrix} \partial\psi/\partial\varphi^\T\\ \partial\lambda/\partial\varphi^\T\end{pmatrix} 
\begin{pmatrix} \partial\varphi/\partial\psi& \partial\varphi/\partial\lambda^\T\end{pmatrix} =
\begin{pmatrix} 
{\partial\psi\over \partial\varphi^\T} {\partial\varphi\over \partial\psi}& {\partial\psi\over \partial\varphi^\T} {\partial\varphi\over \partial\lambda^\T}\\
{\partial\lambda\over \partial\varphi^\T} {\partial\varphi\over \partial\psi}& {\partial\lambda\over \partial\varphi^\T} {\partial\varphi\over \partial\lambda^\T}
\end{pmatrix} , 
\end{equation}
 the first row of the inverse on the right-hand side of~\eqref{inverse.eq} is $\partial \psi/\partial\varphi^\T$, yielding
$$
\hat\psi-\psi = {\partial\psi(\hat\theta_\psi) \over \partial\varphi^\T} \left\{\varphi(\hat\theta) - \varphi(\hat\theta_\psi)\right\} + \cdots. 
$$
Hence $q(\psi)$ should be based on a local departure of ${\hat\theta}$ from $\hat\theta_\psi$, or equivalently $\hat\varphi$ from $\hat\varphi_\psi$, which can be measured through the constructed parameter   $\chi(\theta)$ given by
$$
\chi(\theta) = u^\T \varphi(\theta), \quad u={\partial\psi(\hat\theta_\psi) / \partial\varphi \over \left\| \partial\psi(\hat\theta_\psi)/ \partial\varphi\right\|},
$$
i.e.,  the orthogonal projection of $\varphi(\theta)$ onto a unit vector $u$ parallel to $\partial\psi(\hat\theta_\psi) /\partial\varphi$; note that $\chi(\theta)$ depends on the data through $\hat\theta_\psi$.   The Wald-type measure $q$ from \eqref{r-and-q.eq-q} for $\chi(\theta)$ is 
\begin{equation}
\label{eq.wald1}	
q(\psi) = \sign(\hat\psi-\psi) \left|\chi(\hat\theta)-\chi(\hat\theta_\psi)\right|\left\{{|\jmath(\hat\varphi)|\over |\jmath_{(\lambda\lambda)}(\hat\varphi_\psi)|}\right\}^{1/2},
\end{equation}
where the determinants are computed in the $\varphi$ parametrisation, 
\begin{equation}\label{eq.info} 
\left|\jmath(\hat\varphi)\right|   
= \left|\jmath(\hat\theta)\right| \left| {\partial\varphi(\hat\theta)\over \partial\theta^\T}\right|^{-2}, \\
\quad 
\left|\jmath_{(\lambda\lambda)}(\hat\varphi_\psi)\right| = \left| \jmath_{\lambda\lambda} (\hat\theta_\psi)\right| \left| {\partial\varphi^\T(\hat\theta_\psi)\over \partial\lambda}
{\partial\varphi(\hat\theta_\psi)\over \partial\lambda^\T}\right|^{-1},
\end{equation}
and the second factor on the right-hand side of the second expression here         stems from the ``area formula'' \citep[see, for example,][Lemma~5.1.4]{Krantz.Parks:2008}. 
An equivalent expression for $q(\psi)$ is obtained by substituting these expressions and simplifying, yielding
\begin{equation}\label{q.eq}
q(\psi) = { | \varphi(\hat\theta)-\varphi(\hat\theta_\psi) \quad {\partial\varphi(\hat\theta_\psi)/ \partial \lambda^\T}|  \over 
\left| {\partial\varphi(\hat\theta)/ \partial\theta^\T}\right| }\times \left\{{|\jmath(\hat\theta)|\over |\jmath_{\lambda\lambda}(\hat\theta_\psi)|}\right\}^{1/2}.
\end{equation}
If $\varphi^\T=(\psi,\lambda^\T)$, then~\eqref{q.eq} reduces to~\eqref{r-and-q.eq-q}.  

The equivalence of \eqref{eq.wald1} and \eqref{q.eq} follows by noting that 
the determinant of the $p\times p$ matrix
$$
\begin{pmatrix}  \varphi(\hat\theta)-\varphi(\hat\theta_\psi) & {\partial\varphi(\hat\theta_\psi)/ \partial \lambda^\T}\end{pmatrix} 
$$
is the signed volume $l_1v_2$ of the parallelepiped generated by its $p$ columns, with $l_1$ the length of the component of its first column in the direction orthogonal to its other columns and $v_2$ the volume of the $(p-1)$-dimensional parallelepiped these last columns generate \citep{Fraser.Reid.Wu:1999, Skovgaard:1996}, which is 
$$
v_2 = \left| {\partial\varphi^\T(\hat\theta_\psi)\over \partial\lambda} {\partial\varphi(\hat\theta_\psi)\over \partial\lambda^\T}\right|^{1/2}. 
$$
As the vector $\partial\psi(\hat\theta_\psi) /\partial\varphi$ is orthogonal to ${\partial\varphi(\hat\theta_\psi)/ \partial \lambda^\T}$, we have $l_1= \left|\chi(\hat\theta)-\chi(\hat\theta_\psi)\right|$; the sign of $\hat\psi-\psi$ supplies the sign of $l_1v_2$.

\subsubsection*{Example 2}

To illustrate the development above, consider independent exponential variables $y_1$ and $y_2$ with rate parameters $\lambda\psi$ and $\lambda$; here $\theta=(\psi,\lambda)^\T$.  The corresponding log likelihood is
$$
\logL(\psi,\lambda) = 2\log\lambda + \log\psi - \lambda(\psi y_1 + y_2), \quad \psi, \lambda>0, 
$$
leading to  $\hat\lambda_\psi= 2/(\psi y_1+y_2)$, $\hat\lambda=1/y_2$, $\hat\psi=y_2/y_1$, $|\jmath(\hat\theta)|=1/(\hat\lambda\,\hat\psi)^{2}$ and $|\jmath_{\lambda\lambda}(\hat\theta_\psi)|=2/\hat\lambda_\psi^2$.  

The elements of~\eqref{general-exp-family.eq} are
\begin{eqnarray*}
\varphi&= &\begin{pmatrix}\varphi_1\\ \varphi_2\end{pmatrix} = -\begin{pmatrix} \lambda\psi\\ \lambda\end{pmatrix}, \\
 s(y) &= &\begin{pmatrix} y_1\\ y_2\end{pmatrix},\\ 
 \kappa(\varphi) &= &-\log(-\varphi_1)-\log(-\varphi_2), \quad \varphi_1,\varphi_2<0,
\end{eqnarray*}
so $\psi=\varphi_1/\varphi_2$, and  fixing $\psi$ is equivalent to forcing $\varphi$ to lie on a line through the origin of gradient $1/\psi$.  Now 
$$
{\partial \varphi\over \partial\lambda} = -\begin{pmatrix}\psi\\ 1\end{pmatrix}, \quad {\partial \psi\over \partial\varphi} = \begin{pmatrix}1/\varphi_2\\ -\varphi_1/\varphi_2^{2}\end{pmatrix}
= {1\over \lambda} \begin{pmatrix}-1\\ \psi\end{pmatrix}.
$$
Hence the unit vector in the direction $\partial\psi/\partial\varphi$ is $(1+\psi^2)^{-1/2}(-1,\psi)^\T$, and $u$ is obtained from this by evaluating it at $\varphi(\hat\theta_\psi)$; this is orthogonal to ${\partial \varphi/ \partial\lambda}$ by construction. 

To be concrete, let $y_1=1$ and $y_2=2$, and consider computing the significance function at $\psi=1$.  In this case $\hat\theta_\psi=(1,2/3)^\T$, $\hat\theta=(2,1/2)^\T$, $\varphi(\hat\theta_\psi) = -(2/3,2/3)^\T$, $\varphi(\hat\theta) = -(1,1/2)^\T$ and $u=(-1/\surd{2},1/\surd{2})^\T$.

Figure~\ref{fig2} shows the log likelihoods, the maximum likelihood estimates and the partial maximum likelihood estimates in the $\theta$ and $\varphi$ parametrisations.  The construction of $\chi(\theta) = u^\T \varphi(\theta)$ in terms of $u$ and $\varphi(\theta)$, shown in the right-hand panel, yields $\chi(\hat\theta)=1/\surd{8}$ and $\chi(\hat\theta_\psi)=0$.  The difference $\chi(\hat\theta)-\chi(\hat\theta_\psi)$ in the constructed parameter is a signed distance along the blue dotted line.  As $\psi$ varies, the grey line $\varphi_1=\psi\varphi_2$ and the vector $u$, which is orthogonal to that line, also vary.   Increasing $\psi$ starting from $\psi=1$ would move $\varphi(\hat\theta_\psi)$ along the black dotted line closer to $\varphi(\hat\theta)$, inclining the vector $u$ and the blue dotted line closer to vertical and  reducing significance by decreasing  $\chi(\hat\theta)-\chi(\hat\theta_\psi)$.   When the grey line passes though the black square representing  $\varphi(\hat\theta)$, it coincides with the black circle and the blue square and then $\chi(\hat\theta)-\chi(\hat\theta_\psi)=0$.  Decreasing $\psi$ would move $\varphi(\hat\theta_\psi)$ along the black dotted line away from $\varphi(\hat\theta)$, inclining the vector $u$ and the blue dotted line further from the vertical, and increasing $\chi(\hat\theta)-\chi(\hat\theta_\psi)$ and thus the significance.\endex

\begin{figure}[t]
\begin{center}
\centerline{\includegraphics[width=\textwidth]{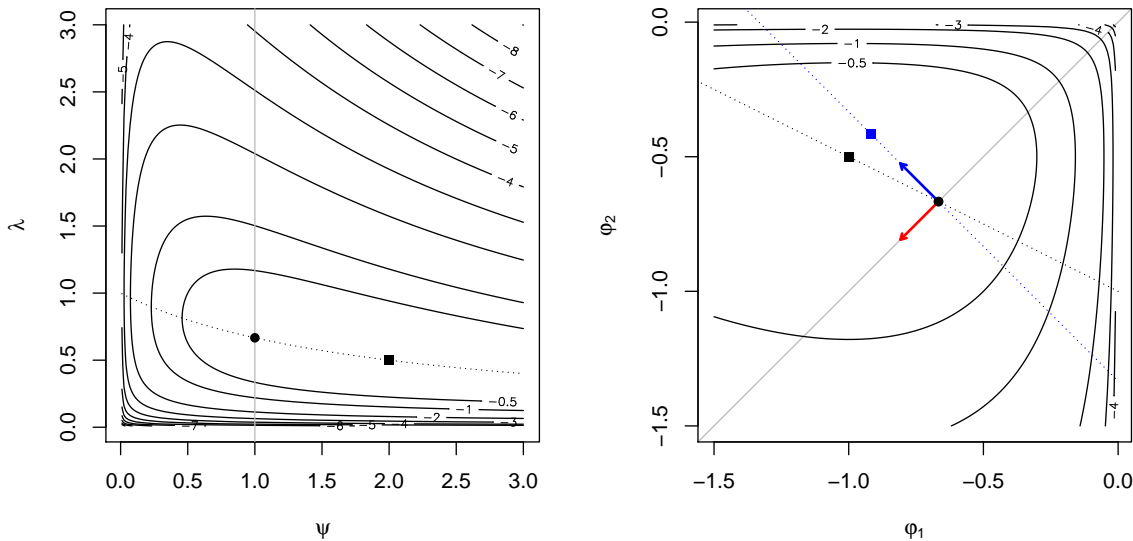}}
\caption{Log likelihoods for  exponential example in $\theta$ parametrisation (left) and  $\varphi$ parametrisation (right).  The solid grey lines show $\psi=1$ or equivalently $\varphi_1=\varphi_2$.  The overall maximum likelihood estimates $\hat\theta$ and $\varphi(\hat\theta)$ are shown by black squares, and the partial maximum likelihood estimates $\hat\theta_\psi$ and $\varphi(\hat\theta_\psi)$, with $\psi=1$, by black circles.  The dotted black lines show $\hat\theta_\psi$ and $\varphi(\hat\theta_\psi)$ as functions of $\psi$.  The arrows in the right-hand panel show the directions of the vectors $\partial\varphi(\hat\theta_\psi)/\partial\lambda$ (red) and $\partial\psi(\hat\theta_\psi)/\partial\varphi$ (blue).  The difference $\chi(\hat\theta)-\chi(\hat\theta_\psi)$ appearing in $q(\psi)$ is a distance along the blue dotted line, between $\chi(\hat\theta_\psi)$ (black filled circle) and $\chi(\hat\theta)$ (blue square). } 
\label{fig2}
\end{center}
\end{figure}

\subsection{Renormalisation}\label{renorm.sect}

Although the error in~\eqref{exp-family.eq4} is ostensibly $O(n^{-1})$, it is actually $O(n^{-3/2})$ in wide generality after renormalisation of the approximation to ensure that it integrates to 1. While in practice this would be done numerically, we note that if the error term in~\eqref{exp-family.eq3} takes the form $B/n$ where $B$ does not depend on the variable of integration, then 
$$
1 = \int f\{r^*(\theta);\theta\}\D{r^*} = \int c\surd(2\pi)\phi\{r^*(\theta)\}\{1 + B/n + O(n^{-3/2})\}\D{r^*}
$$
implies 
$$
c\surd(2\pi) = 1-B/n +O(n^{-3/2}),
$$
so~\eqref{exp-family.eq3} becomes 
$$
f(r^*;\theta) = (1-{B/{n}})\phi\{r^*(\theta)\}\{1+{B/{n}} + O(n^{-3/2})\} = \phi\{r^*(\theta)\}\{1+O(n^{-3/2})\}.
$$
As noted below~\eqref{r-star.eq}, in the  linear exponential family with $p=1$ we indeed have $B$ free of $r^*$. 
In work as yet unpublished Y.~Tang has verified that $B$ is also constant in $r^*$ for the multi-parameter linear exponential family treated in Section~\ref{mult.sec}, and for regression-scale models~\eqref{linear.eq}. 

If the $O(n^{-1})$ error term depends on $r^*(\theta)$, then under mild conditions
$$
B\{r^*(\theta)\} = B(0) + r^*(\theta) B'(s), \quad 0<|s|<|r^*(\theta)|, 
$$
and the term $B(0)/n$ cancels with the normalizing constant as above, so we only need consider
$$
c\surd(2\pi)\phi\{r^*(\theta)\}r^*(\theta)B'(s).
$$
If $B'(s)$ is constant, then this term integrates to $0$ and the norming constant is as before. If $B'(s)$ depends on $r^*$, it would have to to rise very rapidly for the error term to be unbounded, as the normal density function drops rapidly as $|r^*(\theta)|$ increases. Slower, but not constant, dependence of $B'$ on $r^*$ would lead to  relative error just $O(1/n)$, not improved by renormalization. 

Mild regularity conditions \citep{Daniels:1956} ensure the validity of this argument, which also applies to Laplace and similar approximations, including those in Sections~\ref{mult.sec} 
and~\ref{exp-family-general.sect}. A similar analysis verifies that the relative error in the Lugannani--Rice (\citeyear{Lugannani.Rice:1980}) approximation~\eqref{LR.eq} is also $O(n^{-3/2})$.

\section{General likelihood}

\subsection{Basic approximation}\label{basic.sect}

Section~\ref{exp-family:sect} described approximations useful for inference in exponential families.  We now outline how these may be extended to general models, using the tangent exponential model~\eqref{eq.tem}. As noted there, the tangent exponential model has the structure of an exponential family model, with canonical parameter $\varphi$ and cumulant generator $-\ell(\varphi;y^\o)$.  To compare~\eqref{eq.tem} to the general exponential family density~\eqref{general-exp-family.eq}, we take the baseline density $h(s)$ to be $h(s;\hat\theta^\o)$, which effectively centers the score variable $s$ so that $s^\o=0$. Thus we have
\begin{equation}\label{eq.tem.again}
	f_{\rm TEM}(s\mid a;\theta) = \exp\{s^{\T}\varphi(\theta;y^\o) + \ell(\theta;y^{\o})\}h(s;\hat\theta^\o),
\end{equation}
with  canonical variable $s$ and canonical parameter $\varphi(\theta)\equiv \varphi(\theta; y^\o)$ defined locally at $y^\o$, as we now describe.

In an ordinary exponential family the canonical parameter can be obtained (up to linear transformations) from $\partial\ell(\theta;s)/\partial s$. In a general model $\varphi(\theta)$ is also constructed by differentiating in the sample space, now using the $n\times p$ matrix $V$ of sufficient directions of Section~\ref{approx.sect}. As shown there, projecting onto the space spanned by these directions implements conditioning on an approximate ancillary statistic without requiring an explicit form for that ancillary. In more detail,
\begin{equation}
\label{eq.tem3}
	\varphi(\theta;y^\o) = \ell_{;V}(\theta;y^\o),
\end{equation}
where 
$$
\ell_{;V}(\theta;y^\o) =\left.{{d}\over{dt}}\ell(\theta;y^{\o} + Vt)\right|_{t=0}  = V^{\T}\left.{\partial\logL(\theta; y)\over \partial y}\right|_{y=y^\o} 
= \sum_{j=1}^n V_j^{\T}{\partial \logL(\theta; y_j^\o)\over \partial y_j}, 
$$
with $t = (t_1, \dots, t_p)^{\T}$; the last equality applies  when the $y_j$ make independent contributions $\ell(\theta;y_j)$ to the log likelihood. 
Note that both $\ell(\theta;y^\o) = \log f(y^\o;\theta)$ and $\varphi(\theta;y^\o)$ are computed from the original model $f(y;\theta)$.   
 
The validity of this local approximation can be established by Taylor series approximation, as described in Section~\ref{derivation}. Taking \eqref{eq.tem.again} as our starting point, we simply apply the formulae derived for the general exponential family model in Section~\ref{exp-family-general.sect}; computing  $r(\psi)$ at~\eqref{r-and-q.eq-r} and $q(\psi)$ at~\eqref{q.eq}. The approximate significance function for an interest parameter $\psi$ is $\Phi\{r^*(\psi)\}$, using~\eqref{r-star.eq}. 

The tangent model approximates the original density in a neighbourhood of the observed data point $y^\o$; more precisely the original model and the tangent exponential model have the same observed log-likelihood function, and the same sample-space derivative of the log-likelihood function. This turns out to be enough to ensure that the significance function is accurate to $O(n^{-3/2})$. 

The saddlepoint  approximation to  (\ref{eq.tem.again}) is 
\begin{equation}\label{eq.temsaddle}
	f_{\rm TEM}(s  \mid a;\theta )  \doteq c|j(\hat \varphi)|^{-1/2} \exp\bigl[ s^{\T}\{ \varphi(\theta) - \varphi(\hat \theta^\o)\} + \ell(\theta; y^\o) - \ell(\hat \theta^\o; y^\o)\bigr],
\end{equation}
where $j(\varphi)$ is the observed Fisher information computed in the $\varphi$ parametrisation as at \eqref{eq.info}, and $\ell(\theta;y^\o) = \ell\{\theta(\varphi);y^\o\}$. The validity of this saddlepoint approximation is verified by Taylor-series expansion of the log likelihood $\ell(\theta;s)$ about $\hat\theta^\o$ and $s^\o$, described in Section~\ref{derivation}. 

\subsection{Generalising $V$}\label{V.sect}

The matrix $V$ of sufficient directions is central to approximations based on the tangent exponential model.  Our previous discussion has taken $V$ to equal $\partial y/\partial \theta^\T$, evaluated at the observed values $y^\o$  of the data $y$ and the maximum likelihood estimate $\hat\theta^\o$, but this expression presupposes that the individual observations $y_j$ are scalar and can be differentiated with respect to $\theta$; the latter is not the case for discrete responses.  In this section we outline how $V$ can be generalized.

As a preliminary remark, notice that the span of $V$ is invariant to smooth invertible reparametrisation $\theta\mapsto \theta'$, which has the effect of post-multiplying $V$ by an invertible $p\times p$ matrix of constants; see~\eqref{V.eq1}.  It is readily checked that such a multiplication leaves~\eqref{q.eq} unchanged, so  $V$ can be computed in whatever parametrisation is simplest.

When the $y_j$ are continuous and independent vectors of possibly different dimensions $d_j$ we can write 
$$
\varphi(\theta) = \sum_{j=1}^n \left. {\partial y_j^\T\over \partial \theta}\right|_{y_j=y_j^\o,\theta=\hat\theta^\o} \times \left.{\partial \logL(\theta; y_j)\over \partial y_j}\right|_{y_j=y_j^\o} 
=\sum_{j=1}^n V_j^\T \left.{\partial \logL(\theta; y_j)\over \partial y_j}\right|_{y_j=y_j^\o} , 
$$
say, where $V_j$ has dimension $d_j\times p$. This effectively replaces the matrix $V$ by a tensor.

\subsubsection*{Example 3}

If $\{(y_{1j}, y_{2j}), j = 1, \dots, n\}$ are independent pairs from a bivariate normal distribution with zero means, unit variances and covariance $\theta$,  
$V$ can be contructed using the pivotal quantities $z_{1j} = (y_{1j}+y_{2j})^2/\{2(1+\theta)\}$ and $z_{2j}=(y_{1j}-y_{2j})^2/\{2(1-\theta)\}$, leading to
$$
V_j = \left( {{y^\o_{2j}-\hat\theta^\o y^\o_{1j}}\over{1-\hat\theta^{\o 2}}}, {{y^\o_{1j}-\hat\theta^\o y^\o_{2j}}\over{1-\hat\theta^{\o 2}}}\right),\quad j=1,\ldots, n, 
$$
and thus to
$$
\varphi(\theta) = \ell_{;V}(\theta) = {{\theta(t^\o-\hat\theta^\o s^\o)-(s^\o-\hat\theta^\o t^\o)}\over{(1-\theta^2)(1-\hat\theta^{\o 2})}},
$$
where $t=\sum(y_{1j}^2+y_{2j}^2)/2$ and $s=\sum y_{1j}y_{2j}$. The sufficient statistics $(s,t)$ emerge naturally in the construction of $\varphi(\theta)$. If a preliminary reduction to sufficiency is made, the resulting $V$ is a $2\times 1$ vector instead of a $2n \times 1$ vector as above, though $\varphi(\theta)$ is unchanged. The sample space contours determined by $V$ are illustrated in \citet{Reid:2003}, and the accuracy of the normal approximation to the distribution of $r^*$ is illustrated in \citet{Reid.2005}.
\endex

Computing $V$ for discrete responses is more awkward.  In Section~\ref{computation:sect} we saw that in the continuous case, total differentiation of the pivot $F_j(y_j;\theta)$ led to the expression~\eqref{altV.eq} for $V_j$, but $\partial F_j(y;\theta)/\partial y=0$ almost everywhere in the discrete case.  To deal with this, note that in a continuous exponential family model with canonical observation $y_j$, we can write $\partial \logL(\theta; y_j)/ \partial y_j =\alpha_j(\theta)$, say, and observe that as
\begin{eqnarray*}
n^{-1}\varphi(\theta) &=&n^{-1}\sum_{j=1}^n  {\partial y_j^\T\over \partial\theta} \alpha_j(\theta) \\
&=& n^{-1}\sum_{j=1}^n  \E\left({\partial y_j^\T\over \partial\theta}\right) \alpha_j(\theta)
+ n^{-1}\sum_{j=1}^n \left\{ {\partial y_j^\T\over \partial\theta} - \E\left({\partial y_j^\T\over \partial\theta}\right) \right\} \alpha_j(\theta) ,
\end{eqnarray*}
and each term in the final sum has mean zero, that term is $O_p(n^{-1/2})$.  Thus if the order of integration and differentiation can be interchanged, $n^{-1}\varphi(\theta) $ can be replaced with
$$
n^{-1}\tilde\varphi(\theta) = n^{-1}\sum_{j=1}^n \left.{\partial \E(y^\T_j)\over \partial\theta}\right|_{\theta=\hat\theta^\o}  \alpha_j(\theta),
$$
at the expense of introducing an  $O_p(n^{-1/2})$ error.  Hence using $\tilde\varphi(\theta)$ in~\eqref{q.eq} does not change the $O(n^{-1})$ error in~\eqref{exp-family.eq4}.  The use of $\partial \E(y^\T_j)/ \partial\theta$ replaces the sufficient directions  with their expectation, which is only  tangential to $\calA^\o$ on average, so renormalisation no longer reduces the order of error to $n^{-3/2}$: inference accurate to third order is unavailable.  

As the expectation of a discrete response is typically continuous in the parameters, this approach can be applied to discrete exponential family models such as the Poisson, binomial and multinomial.  Extension to more general discrete response distributions entails replacing $\alpha_j(\theta)$.  \citet{Davison.Fraser.Reid:2006} show that one can use a locally-defined score variable
$w_j$, and
$$
w_j = \left.{\partial \logL(\theta; y_j)\over \partial \theta}\right|_{\theta=\hat\theta^\o}, \quad V_j = \left.{\partial \E(w_j;\theta) \over \partial \theta^\T}\right|_{\theta=\hat\theta^\o}, 
\quad 
\varphi(\theta) = \sum_{j=1}^n  V^\T_j {\partial \logL(\theta; y_j)\over \partial w_j}.
$$
Here $w_j$  has dimension $p\times 1$, so $V_j$ is a $p\times p$ matrix that is easily seen to be the contribution from $y_j$ to the expected information matrix, evaluated at $\hat\theta^\o$.   The derivative $\partial \logL(\theta; y_j)/\partial w_j$ is most easily computed as $\partial \logL(\theta; y_j)/\partial y_j \times ( \partial w_j/\partial y_j)^{-1}$.

\citet{Skovgaard:1996} derived a  version of $r^*$ that replaces $q$ in~\eqref{r-star.eq} with a quantity that is computed entirely from cumulants of the log likelihood. The resulting approximation has a relative error that is $O(n^{-1})$ in a large deviation region about the maximum likelihood estimator.  This provides highly accurate results far out into the tails of its distribution, which Skovgaard argues may be of more practical value than higher, $O(n^{-3/2})$, accuracy near its mean. \citet{Reid.Fraser:2010} show that Skovgaard's approximation can be related to a tangent exponential model with canonical parameter determined from the derivative of $I(\theta;\hat\theta^\o) = \int \ell(\theta;y)f(y;\hat\theta^\o)dy$.  It can also be used for discrete models, and gives the same approximation as \citet{Davison.Fraser.Reid:2006} in curved exponential families, but not more generally.

\section{Derivation of the tangent exponential model}\label{derivation}

\subsection{Preliminary remarks}

The expression for the tangent exponential model,~\eqref{eq.tem}, is concise and emphasizes the connection to exponential family models and the role of $\varphi$ as a canonical parameter, but does not lend itself to ready understanding.  It can be derived using Taylor series approximations that can  be given explicitly when $y$ and $\theta$ are scalar and provide some theoretical illumination.   The development for higher dimensions is similar but much more laborious and does not yield additional insights.
 
We have seen above that  the tangent exponential model requires computation of a first derivative in the sample space.  We can think of this as assessing how the log likelihood $\ell(\theta;y^\o)$ changes not only as a function of $\theta$, as is standard in both likelihood and Bayesian inference, but also as a function of $y$, in a small neighbourhood of the observed data point $y^\o$.  Employing this first derivative probes $\ell(\theta;y)$ more deeply than simply using the observed log likelihood function $\ell(\theta;y^\o)$, but does not involve computing the log likelihood function on its entire domain, i.e., for  $(\theta, y)\in \mathbb{R}^p\times\mathbb{R}^n$.

\subsection{No nuisance parameters}\label{nonuisance}

Suppose that we have a model $f(s;\theta)$, $s\in\mathbb{R}$ and $\theta\in\mathbb{R}$, and that there is an implicit dependence on $n$, in the sense that $\ell(\theta;s) = \log f(s;\theta)$ is $O_p(n)$.   We will address how to get this reduction in general at the end of  this section, but this would be the case for example if $s$ was the sufficient statistic based on a random sample from a linear exponential family model, and it is also the case for models like the Cauchy, where $s$ is the maximum likelihood estimator or any other location-equivariant estimator of $\theta$, based on a sample of size $n$, and the distribution is conditional on the {$(n-1)$-dimensional} ancillary statistic $a$.

We first expand the log likelihood $\log f(s;\theta)$ in a Taylor series in both $s$ and $\theta$, about the fixed points $s^\o$ and $\hat\theta^\o$, giving
\begin{eqnarray}
\nonumber \ell(\theta;s) &=& \log f(s;\theta)\\
\nonumber & =& \ell(\hat\theta^\o; s^\o) + (s - s^\o)\ell_{s}(\hat\theta^\o; s^\o) + (\theta - \hat\theta^\o)\ell_\theta(\hat\theta^\o; s^\o) \\
\nonumber && + \frac{1}{2}(s-s^\o)^2\ell_{ss}(\hat\theta^\o; s^\o) + (s - s^\o)(\theta-\hat\theta^\o)\ell_{\theta s}(\hat\theta^\o; s^\o) \\
\nonumber && + \frac{1}{2}(\theta-\hat\theta^\o)^2\ell_{\theta\theta}(\hat\theta^\o; s^\o) + \cdots \\
&=&\sum_{i,j=0}^\infty\dfrac{1}{i!j!} 
(\theta - \hat\theta^\o)^i(s-s^\o)^j  
b_{ij},\label{exp-expansion.eq}
\end{eqnarray}
say, where {$b_{ij} = \partial^i\partial^j \ell(\hat\theta^\o;s^o)/\partial\theta^i\partial s^j$} for $ i ,j= 0, 1, \dots$;  note that $b_{10}=0$.

Now consider how expansion~\eqref{exp-expansion.eq} would differ if $f(s;\theta)$ were an exponential family model with canonical parameter $\theta$ and sufficient statistic $s$.  In this case
$$
\log f(s;\theta) = \ell(\theta;s) = \theta s - \kappa(\theta) - d(s), 
$$
so the coefficients $b_{i0}$ would be those for a Taylor series expansion of $-d(s)$,  the $b_{0j}$ would be those for a Taylor series expansion of $-\kappa(\theta)$, and the only other non-zero term would be $b_{11} = 1$.

\citet{Andrews.Fraser.Wong:2005} show  that for any continuously differentiable model $f(s;\theta)$ there exists a transformation $x=x(s)$ and $\varphi = \varphi(\theta)$ such that the expansion of $\log f(x;\varphi)$ has the coefficient array starting with $i=j=0$ at the top left and terms $b_{ij}$ in row $i$ and column $j$, given by
 
{\footnotesize \begin{equation}\label{array}
\left(
\begin{array}{ccccc}
b+\dfrac{3\alpha_4-5\alpha_3^2-12\gamma}{24n} & \dfrac{-\alpha_3}{2n^{1/2}}& - \left(1+\dfrac{\alpha_4-2\alpha_3^2-5\gamma}{2n}\right) & \dfrac{\alpha_3}{n^{1/2}} & \dfrac{\alpha_4-3\alpha_3^2-6\gamma}{n} \\
0 & 1 & 0 & 0 &  \\
-1 &0 & \dfrac{\gamma}{n} & & \\
-\dfrac{\alpha_3}{n^{1/2}} & 0 & & &  \\
-\dfrac{\alpha_4}{n} &  & &  & 
\end{array}
\right ).
\end{equation}}
Only terms up to $O(n^{-1})$ are shown: the terms in the blank spaces are $O(n^{-3/2})$ or smaller, as are the terms implicitly omitted.    The constants $\alpha_3$, $\alpha_4$ and $\gamma$ are the derivatives of $\log f(x;\varphi)$ at $x=0$ and $\varphi = 0$:
\begin{eqnarray*}
\alpha_3 &=& -\left.\dfrac{\partial^3\log f(x;\varphi)}{\partial \varphi^3}\right|_{x=0,\varphi=0}, \\
\alpha_4 &=& -\left.\dfrac{\partial^4\log f(x;\varphi)}{\partial \varphi^4}\right|_{x=0,\varphi=0}, \\
\gamma &=& \left.\dfrac{\partial^4 \log f(x;\varphi)}{\partial x^2 \partial\varphi^2}\right|_{x=0,\varphi=0},
\end{eqnarray*}
and $\gamma$ is related to the exponential curvature of the model \citep{Efron:1975}.  Ignoring terms of  $O(n^{-3/2})$, the expansion~\eqref{array} in terms of $x$ and $\varphi$ is almost that of an exponential family model; the only additional coefficient is the $(2,2)$ entry $\gamma/n$, which adds a term $\gamma\varphi^2x^2/(4n)$ to the log likelihood expansion.  

The variables $s$ and $\theta$ are both scaled and centered as part of the transformation to $x$ and $\varphi$: note that the observed information $-\partial^2\log f(0;0)/\partial\varphi^2=-b_{20}=1$. The point $(x,\varphi) = (0,0)$ corresponds to the original point of expansion $(s^\o, \hat\theta^\o)$, where $s^\o$ is the observed value and $\hat\theta^\o = \hat\theta(s^\o)$ is the corresponding value of the maximum likelihood estimator.

Another way to write the model given by~\eqref{array} is 
$$
\log f(x;\varphi) = b_{00} + P_{1n}(x) + P_{2n}(\varphi) + x\varphi + \gamma x^2\varphi^2/(4n) + O(n^{-3/2}),
$$
where $P_{1n}(x)$ is given by the first row of the array and $P_{2n}(\varphi)$ by its first column, each omitting $b_{00}$. On examining the elements of~\eqref{array}  we see that we can write 
\begin{equation}\label{f.eq}
f(x;\varphi) \propto \phi(x-\varphi)\left\{ 1 + {a_1(x,\varphi)\over n^{1/2}} +  {a_2(x,\varphi)\over n}+\gamma {a_3(x,\varphi)\over n}+O(n^{-3/2}) \right\},
\end{equation}
in terms of the standard normal density function $\phi$ and suitable polynomials $a_1$, $a_2$ and $a_3(x,\varphi) = (x^2\varphi^2 -x^4 + 5x^2 -2)/4$.  Equation~\eqref{f.eq} can be integrated term by term with respect to $x$; an explicit array for the resulting approximate distribution function $F(x;\varphi)$ is given in \citet{Andrews.Fraser.Wong:2005}.    Remarkably, although $F(x;\varphi)$ depends on $\gamma$ in general,  $F(0;\varphi)$ does not depend on $\gamma$, because $\phi(x-\varphi)a_3(x;\varphi)$ has integral zero over the negative half-line.  As $x=0$ corresponds to $s=s^\o$, the significance function $F(s^\o;\theta)$ does not depend on $\gamma$ and can be computed using the exponential family version of~\eqref{array}, in which $\gamma=0$.

The tangent exponential model~\eqref{eq.tem} is just an invariant version of the simplified expansion~\eqref{f.eq}: $s$ is now the transformed variable called $x$ here, and is the corresponding score variable, and $\varphi$ is by definition $\partial \log f_{\rm TEM}(0;0)/\partial x$, the canonical parameter of the exponential model approximation.  

The steps in going from the original model to~\eqref{array} are outlined in  \citet{Andrews.Fraser.Wong:2005} and \citet{Cakmak:1998}.

The reduction to a single variable $s$ in a scalar parameter model is straightforward if, for example,  $y = (y_1, \dots, y_n)$ is a sample from an exponential family model, with density function~\eqref{exp-family.eq0}, as the log likelihood $\ell(\theta;s) $ then equals $ \exp\{\varphi(\theta)s - nc(\theta)\}$, where $s = \sum_{i=1}^n s(y_i)$, and has the dependence on $n$ summarized in~\eqref{array}, with $\gamma=0$.

Similarly, in the case of a sample $y = (y_1, \dots, y_n)$ from a location model, the exact distribution of any location-invariant estimator, say $s$, of the location parameter $\theta$ given the location ancillary statistic $a = (y_1-s, \dots, y_n-s)$ is 
$$
f(s\mid a;\theta) = \exp\{\ell(\theta;s,a)\}\left/\int_t\exp \{\ell(t;s,a)\}\,\D{t},\right. 
$$
and~\eqref{array} is equivalent to the density approximation arising when Laplace's method is applied to the denominator integral.

For more general models, the discussion in Section~\ref{approx.sect} establishes the existence of a conditional distribution on $\mathbb{R}$ that can be determined by finding the $n \times 1$ vector $V$ of sufficient directions. The arguments above show that this conditional distribution, which now has a scalar variable and scalar parameter, is effectively an exponential family for the purpose of approximating the significance function.%

\subsection{Nuisance parameters}\label{aargh}

The expansion in~\eqref{array} can be generalized to vector parameters, as in 
\cite{Cakmak.Fraser.Reid:1994} and \cite{Fraser.Reid:1993}, but the notation is  cumbersome, and the various multi-dimensional analogues to $\alpha_3, \alpha_4, \gamma$ are not explicitly available. However, the expansion verifies that the coordinate-free version of the tangent exponential model has the form given at~\eqref{eq.tem} and~\eqref{eq.tem.again}, with saddlepoint approximation~\eqref{eq.temsaddle}.

This gives a tangent exponential model on $\mathbb{R}^p$ for inference about $\theta$, which is implicitly conditioned on an approximate ancillary statistic through the use of the $n \times p$ matrix $V$, and this is now the full model used to obtain an approximate significance function for a scalar parameter of interest $\psi$. 

In this full model, consider fixing $\psi$, and constructing a new tangent exponential model on $\mathbb{R}^{p-1}$ with parameter $\lambda$. We can write, suppressing the conditioning on $a$,
\begin{equation}\label{eq.tem2}
f_{\rm TEM}(s;\theta) = f_{1, \rm TEM}(s_{\psi}\mid \tilde a_\psi;\lambda)f_2(\tilde a_\psi),	
\end{equation}
 where $\tilde a_\psi$ is a new approximate ancillary statistic for the model with $\psi$ held fixed. This gives us a one-dimensional distribution for inference about $\psi$, 
 \begin{equation}\nonumber 
 f_2(\tilde a_\psi) = 	f_{\rm TEM}(s;\theta)\slash f_{1,\rm TEM}(s_{\psi}\mid \tilde a_\psi;\lambda),
 \end{equation}
and as we know the left-hand side is free of both $s_{\psi}$ and $\lambda$, we can choose $s_{\psi}=0$ and $\lambda = \hat\lambda_\psi$. Using the saddlepoint form~\eqref{eq.temsaddle} of the tangent exponential model in the numerator and denominator yields a model on $\mathbb{R}$ of the form 
\begin{equation}\label{eq.tem4}
	h(s;\psi) = c\exp\{\ell(\hat\varphi_\psi;s)-\ell(\hat\varphi;s)\}|\jmath_{\varphi\varphi}(\hat\varphi)|^{-1/2}|\jmath_{(\lambda\lambda)}(\hat\varphi_\psi)|^{1/2}, \quad s \in {\cal L}_\psi,
\end{equation}
where $\ell(\varphi;s) = s^{\T}\varphi + \ell(\varphi;y^\o)$, and ${\cal L}_\psi$ is a line in the sample space corresponding to fixing $\hat\lambda_\psi$ (and $a$). Expressing the result with a constraint on $s$ avoids explicit identification of $\tilde a_\psi$: it is enough to know that it exists.  As in Section~\ref{exp-family-general.sect} the information determinants are computed in the $\varphi$ parameterization; see~\eqref{eq.info}.  

The right-hand-side of~\eqref{eq.tem4}  has the form of our original tangent exponential model~\eqref{eq.temsaddle}, with an adjustment factor in the ratio of determinants; note also the similarity to the approximate conditional density~\eqref{expcond.eq} for linear exponential families.
As a result, the approximate significance function is the same as that for general exponential families outlined in Section~\ref{exp-family-general.sect}, with the significance function as in~\eqref{exp-family.eq4} or~\eqref{LR.eq}, with $r$ defined in~\eqref{r-and-q.eq-r}, and $q$ defined in~\eqref{q.eq}. Once the tangent exponential approximation to the original model has been established, the exponential model formulas apply directly.

It would be natural to partition $s$ into a component related to $\psi$ and one related to $\lambda$, and this is how the result is presented in \citet[Sec.~6]{Fraser.Reid:1995}. In later work \citep{Reid.Fraser:2010, Fraser.Reid.Sartori:2016} the simpler notation of \eqref{eq.tem4} is preferred, with a constraint on $s$  to emphasize that the density is for a variable of the same dimension as the parameter of interest $\psi$.

\section{Concluding remarks}
\label{disc.sect} 

\subsection{Summary}\label{summ.sect}

The tangent exponential model and associated significance function implement inference conditional on an approximate ancillary statistic, followed by marginalization to a pivotal quantity, $r^*(\psi;y)$, for a scalar parameter of interest. This pivot is readily computed using only $\ell(\theta;y^\o)$ and $\varphi(\theta;y^\o)=\ell_{;V}(\theta;y^\o)$, and the full and constrained maximum likelihood estimators $\hat\theta$ and $\hat\theta_\psi$. \citet{Fraser.Reid.Wu:1999} and \citet{Reid:2003} present this ``inference algorithm'' as two dimension-reduction steps: from a model $f(y;\theta)$ on $\mathbb{R}^n$ to a model $f(s\mid a;\theta)$ on $\mathbb{R}^p$, by conditioning, and from this model  to another on $\mathbb{R}$, by marginalizing.  The model on $\mathbb{R}$ can be approximated by a simple standard normal distribution for the pivotal quantity $r^*(\psi;y)$, and in continuous models the approximation to the significance function based on $r^*(\psi;y^\o)$ has relative error $O(n^{-3/2})$. 

The final approximation step is somewhat separate from the development of the model on $\mathbb{R}$, and follows closely the derivation of the $r^*$ approximation in \citet{BN1986}. It can also be applied in other contexts, and in particular to approximation of a Bayesian posterior survivor function, starting from the the Laplace approximation to the posterior marginal density~\citep{Tierney.Kadane:1986}. 

As our focus here is on the steps leading to the tangent exponential model and their implications for inference, we have not included numerical work indicating the accuracy of the approximations. There are many examples and exercises in \citet{Brazzale.Davison.Reid:2007}, in the literature referred to there and in \citet{Brazzale.Davison:2008}. 

There is a close relation between the $r^*$ approximation to the parametric bootstrap; the higher-order properties of the latter are investigated in \citet{DiCiccio.Young:2008}, \citet{Lee.Young:2005}, and \citet{DiCiccioetal:2015}. To achieve the same order of accuracy it is necessary to bootstrap under the constrained maximum likelihood estimate $(\psi, \hat\lambda_\psi)$, which increases the computational burden. \citet{Fraser.Rousseau:2008} also consider the relationship between significance functions based on the parametric bootstrap, on $r^*$, and on Bayesian versions of predictive $p$-values.

\subsection{Extensions}\label{ext.sect}

If the parameter of interest is a vector, a significance function is not easily obtained unless one can construct a scalar measure of departure such as the log likelihood ratio statistic $w(\psi) = 2\{\ell(\hat\theta)-\ell(\hat\theta_\psi)\}$, Wald statistic $(\hat\psi - \psi)^{\T}j^{\psi\psi}(\hat\theta)(\hat\psi-\psi)$, or score statistic, each to first order approximately distributed as $\chi^2_d$. \citet{Davison.Fraser.Reid.Sartori:2014} and \citet{Fraser.Reid.Sartori:2016} use the tangent exponential model as the building block for a directional approach to inference for a $d$-dimensional parameter $\psi$ which creates a univariate summary, by considering the magnitude of $\psi$ conditional on its direction from a null value $\psi_0$. The saddlepoint approximation to $f_{\rm TEM}(s\mid a;\theta)$ on this line in the sample space forms the basis for inference. A new scalar-parameter exponential family is constructed from the multi-parameter exponential family model or the approximating tangent exponential model. 
 
 The discussion above has presumed that the underlying data are independent, but the geometric motivation in Section~\ref{approx.sect} suggests that the approach should provide improved accuracy more generally.   \citet{Belzile.Davison:2021} adapt the approach for discrete responses to the inhomogeneous Poisson process, but this is a special case owing to its independence properties.  The main difficulty in broader settings is to compute $V$, and from this the constructed parameter $\varphi(\theta)$.  In a time series setting, a series of pivotal quantities may be generated from the predictive distributions $F(y_j\mid y_{j-1},\ldots, y_1;\theta)$ for $ j = 2, \dots, n$, using martingale differences or a lower triangular square root of the covariance matrix for the response  \citep{Fraser.Rekkas.Wong:2005,Lozada-Can.Davison:2010}. It is not yet clear whether other decompositions of the covariance matrix would lead to asymptotically equivalent results.  

There is an $r^*$ approximation for Bayesian inference, readily obtained from the Laplace approximation, as mentioned above; see also \citet{Fraser.Reid.Wu:1999}. This provides a route to examining the discrepancy between posterior survivor functions and significance functions. Equating the two versions of $r^*$ leads to a data-dependent prior that ensures agreement of the significance and survivor functions up to terms of $O(n^{-1})$. The former was emphasised in \cite{Fraser:2011} and \citet{Fraser.etal:2016}; the latter formed the basis for a discussion of default priors in \cite{FMRY:2010}.

\subsection{A brief historical note}\label{hist.sect}

Fraser viewed the dimension-reduction steps in Section~\ref{summ.sect} as essentially unique, and consequently the pivotal quantity $r_\psi^*$ not as an arbitrary choice among several potential pivotal quantities, but as the only route to higher-order approximation for a scalar parameter in the presence of nuisance parameters: 
\begin{quote}
This ancillary density is uniquely determined by steps that retain continuity of the model 	in the derivation of the marginal distribution. It thus provides the unique null density for assessing a value $\psi=\psi_0$, and anyone suggesting a different null distribution would need to justify inserting discontinuity where none was present \citep[\S 4]{Fraser:2017}.
\end{quote}
The continuity referred to there is the presumption that changes in $\theta$ are smoothly related to changes in $y$ and vice-versa, as in a pure location model $f(y-\theta)$. The vectors $V$ determining the tangent plane to the ancillary surface are based on the local location model defined in Fraser (1964). Suppose $y_i$ has density $f(y_i;\theta)$ and cumulative distribution function $F(y_i;\theta), \theta\in\mathbb{R}$. Define a transformation $y_i\mapsto x_i$ by setting
$$
 x_i = \int^{y_i}\ - {{F_{y}(y;\theta_0)}\over{F_{\theta}(y;\theta_0)}}\,\D{y}, 
$$
where $\theta_0$ is some fixed value. The density of $x_i$ has location model form near $\theta_0$, and this local location model has an ancillary statistic $(x_1-\bar x, \dots, x_n-\bar x)$, and sufficient direction $(1, \dots, 1)$, which {transforms back to the sufficient direction
$$   
V^{\T} =   - \left({{F_{\theta}(y_1;\theta_0)}\over{F_{y}(y_1;\theta_0)}} , \dots, {{F_{\theta}(y_n;\theta_0)}\over{F_{y}(y_n;\theta_0)}}\right)
$$
in terms of $y_1,\ldots, y_n$; see~\eqref{altV.eq}. As noted in \citet{Fraser.Reid:2001},  this construction does not give a local location model for the full sample $y_1, \dots, y_n$ because the vector field $V(y)$ is not guaranteed to be integrable. But the expansions in that paper verify that the approximations derived from the tangent exponential model are still valid, as the sufficient directions $V$ describe the same tangent plane as a second-order ancillary statistic that exists under mild regularity conditions. \citet[\S 3]{Fraser.Reid:2001} promised that ``the integrability of the $V(y)$ to the required order will be examined elsewhere'', and this was fulfilled in \citet{Fraser.Fraser.Staicu:2010}.
 
Fraser viewed as intrinsically linked the construction of the tangent exponential model, the application of the saddlepoint approximation and the construction of significance functions as key inferential summaries. A first, lengthy, paper written shortly after the simpler developments in \cite{Fraser:1988, Fraser:1990, Fraser:1991} included all these pieces, and was met with some puzzlement by editors and reviewers: one reviewer advised ``it should probably be several papers'' --- a reaction that might be rather unusual nowadays. This led to the asymptotic expansions being the focus of \cite{Fraser.Reid:1993}, although much of the original draft was published in \cite{Fraser.Reid:1995}. That latter paper derived the tangent exponential approximation to general models, derived the directional vectors $V$ from a local location model, showed the existence of a second-order ancillary statistic with the same directional vectors, verified that the dimension of this ancillary is fixed as $n\rightarrow\infty$, and derived the $r^*$ approximation in its general form. The construction of the directional vectors was discussed in more detail in \cite{Fraser.Reid:2001}, which is confusingly referred to in some of his papers as Fraser and Reid (1999). 

The annotations in the bibliography below attempt to provide a road map through the most relevant of these papers. Copies of the less readily accessible ones are posted at 
\begin{verbatim}
https://utstat.toronto.edu/reid/fraser-papers.html
\end{verbatim}
 
\subsubsection*{Acknowledgements}

The work was supported by the Swiss National Science Foundation and the Natural Sciences and Engineering Council of Canada. We thank L\'eo Belzile, Yanbo Tang and two anonymous referees for helpful comments. 

\bibliographystyle{input-files/CUP}

\bibliography{input-files/mybib}

\end{document}